\documentclass[twocolumn]{aastex63}

\pdfoutput=1 
\usepackage[figure,figure*]{hypcap}
\usepackage{footnote}
\usepackage{hyperref}
\usepackage{subfigure}
\usepackage{amsmath}

\usepackage{epsfig}
\usepackage{natbib}

\shorttitle{3-D CMZ I}
\shortauthors{Battersby et al.}

\begin{document}

\def\Msun{\hbox{M$_{\odot}$}}
\def\Lsun{\hbox{L$_{\odot}$}}
\def\kms{km~s$^{\rm -1}$}
\def\hcop{HCO$^{+}$}
\def\n2hp{N$_{2}$H$^{+}$}
\def\micron{$\mu$m}
\def\13CO{$^{13}$CO}
\def\etamb{$\eta_{\rm mb}$}
\def\Inu{I$_{\nu}$}
\def\kapnu{$\kappa _{\nu}$}
\def\ffore{f$_{\rm{fore}}$}
\def\tastar{T$_{A}^{*}$}
\def\nh3{NH$_{3}$}
\def\deg{$^{\circ}$}
\def\arcsec{$^{\prime\prime}$}
\def\arcmin{$^{\prime}$}
\def\Vlsr{\hbox{V$_{LSR}$}}
\def\newv2{\textcolor{orange}}

\title{3-D CMZ I: Central Molecular Zone Overview} 

\author[0000-0002-6073-9320]{Cara Battersby}
\affiliation{University of Connecticut, Department of Physics, 196A Hillside Road, Unit 3046
Storrs, CT 06269-3046, USA}
\affiliation{Center for Astrophysics $|$ Harvard \& Smithsonian, MS-78, 60 Garden St., Cambridge, MA 02138 USA}
\author[0000-0001-7330-8856]{Daniel L. Walker}
\affiliation{UK ALMA Regional Centre Node, Jodrell Bank Centre for Astrophysics, Oxford Road, The University of Manchester, Manchester M13 9PL, United Kingdom}
\affiliation{University of Connecticut, Department of Physics, 196A Hillside Road, Unit 3046
Storrs, CT 06269-3046, USA}

\author[0000-0003-0410-4504]{Ashley Barnes}
\affiliation{European Southern Observatory (ESO), Karl-Schwarzschild-Straße 2, 85748 Garching, Germany}MA

\author[0000-0001-6431-9633]{Adam Ginsburg}
\affiliation{University of Florida Department of Astronomy, Bryant Space Science Center, Gainesville, FL, 32611, USA}

\author[0000-0002-5776-9473]{Dani Lipman}
\affiliation{University of Connecticut, Department of Physics, 196A Hillside Road, Unit 3046
Storrs, CT 06269-3046, USA}

\author[0009-0005-9578-2192]{Danya Alboslani}
\affiliation{University of Connecticut, Department of Physics, 196A Hillside Road, Unit 3046
Storrs, CT 06269-3046, USA}

\author[0000-0003-0946-4365]{H Perry Hatchfield}
\affiliation{{Jet Propulsion Laboratory, California Institute of Technology, 4800 Oak Grove Drive, Pasadena, CA, 91109, USA}}
\affiliation{University of Connecticut, Department of Physics, 196A Hillside Road, Unit 3046
Storrs, CT 06269-3046, USA}

\author[0000-0001-8135-6612]{John Bally}
\affiliation{CASA, University of Colorado, 389-UCB, Boulder, CO 80309}

\author[0000-0001-6708-1317]{Simon C.~O.\ Glover}
\affiliation{Universit\"{a}t Heidelberg, Zentrum f\"{u}r Astronomie, Institut f\"{u}r Theoretische Astrophysik, Albert-Ueberle-Str.\ 2, 69120 Heidelberg, Germany}

\author[0000-0001-9656-7682]{Jonathan~D.~Henshaw}
\affiliation{Astrophysics Research Institute, Liverpool John Moores University, 146 Brownlow Hill, Liverpool L3 5RF, UK}
\affiliation{Max-Planck-Institut f\"ur Astronomie, K\"onigstuhl 17, D-69117 Heidelberg, Germany}

\author[0000-0003-4140-5138]{Katharina Immer}
\affiliation{European Southern Observatory (ESO), Karl-Schwarzschild-Straße 2, 85748 Garching, Germany}MA

\author[0000-0002-0560-3172]{Ralf S.\ Klessen}
\affiliation{Universit\"{a}t Heidelberg, Zentrum f\"{u}r Astronomie, Institut f\"{u}r Theoretische Astrophysik, Albert-Ueberle-Str.\ 2, 69120 Heidelberg, Germany}
\affiliation{Universit\"{a}t Heidelberg, Interdisziplin\"{a}res Zentrum f\"{u}r Wissenschaftliches Rechnen, Im Neuenheimer Feld 225, 69120 Heidelberg, Germany}
\affiliation{Center for Astrophysics $|$ Harvard \& Smithsonian, MS-78, 60 Garden St., Cambridge, MA 02138 USA}
\affiliation{Radcliffe Institute for Advanced Studies at Harvard University, 10 Garden Street, Cambridge, MA 02138, USA}

\author[0000-0001-6353-0170]{Steven N.~Longmore}
\affiliation{Astrophysics Research Institute, Liverpool John Moores University, 146 Brownlow Hill, Liverpool L3 5RF, UK}

\author[0000-0001-8782-1992]{Elisabeth A.~C.~Mills}
\affiliation{Department of Physics and Astronomy, University of Kansas, 1251 Wescoe Hall Drive, Lawrence, KS 66045, USA}

\author{Sergio Molinari}
\affiliation{INAF - Istituto di Astrofisica e Planetologia Spaziale, Via Fosso del Cavaliere 100, I-00133 Roma, Italy}

\author[0000-0002-0820-1814]{Rowan Smith}
\affiliation{Scottish Universities Physics Alliance (SUPA), School of Physics and Astronomy, University of St Andrews, North Haugh, St Andrews KY16 9SS, UK}
\affiliation{Jodrell Bank Centre for Astrophysics, Department of Physics and Astronomy, University of Manchester, Oxford Road, Manchester M13 9PL, UK}

\author[0000-0001-6113-6241]{Mattia C.\ Sormani}
\affiliation{Universit{\`a} dell’Insubria, via Valleggio 11, 22100 Como, Italy}

\author[0000-0002-9483-7164]{Robin~G.~Tress}
\affiliation{Institute of Physics, Laboratory for Galaxy Evolution and Spectral Modelling, EPFL, Observatoire de Sauverny, Chemin Pegasi 51, 1290 Versoix, Switzerland}

\author[0000-0003-2384-6589]{Qizhou Zhang}
\affiliation{Center for Astrophysics $|$ Harvard \& Smithsonian, MS-78, 60 Garden St., Cambridge, MA 02138 USA}

\begin{abstract}
The Central Molecular Zone (CMZ) is the largest reservoir of dense molecular gas in the Galaxy and is heavily obscured in the optical and near-IR. We present an overview of the far-IR dust continuum, where the molecular clouds are revealed, provided by Herschel in the inner 40\deg~($|l| <$ 20\deg) of the Milky Way with a particular focus on the CMZ. We report a total dense gas ($N$(H$_2$) $> 10^{23}$ cm$^{-2}$) CMZ mass of  $\sim 2\substack{+2 \\ -1} \times 10^7$ \Msun~and confirm that there is a highly asymmetric distribution of dense gas, with about 70-75\% at positive longitudes. We create and publicly release complete fore/background-subtracted column density and dust temperature maps in the inner 40\deg~($|l| <$ 20\deg) of the Galaxy. 
We find that the CMZ clearly stands out as a distinct structure, with an average mass per longitude that is at least $3\times$ higher than the rest of the inner Galaxy contiguously from 1.8\deg $> \ell >$ -1.3\deg. This CMZ extent is larger than previously assumed, but is consistent with constraints from velocity information. 
The inner Galaxy's column density peaks towards the SgrB2 complex with a value of about 2 $\times$ 10$^{24}$  cm$^{-2}$, and typical CMZ molecular clouds are about N(H$_2$) $\sim$10$^{23}$ cm$^{-2}$. Typical CMZ dust temperatures range from about $12-35$ K with relatively little variation. We identify a ridge of warm dust in the inner CMZ that potentially traces the base of the northern Galactic outflow seen with MEERKAT.
\end{abstract}

\section{Introduction}
\label{sec:intro}

Studies of our Milky Way Galaxy afford unparalleled resolution and allow for detailed study of the gas flows, star formation, and feedback processes that drive galaxies. However, it is challenging to gain a large-scale perspective of the structure and dynamics of the Galaxy from our location embedded within its mid-plane. The last few decades have provided numerous comprehensive surveys of the plane of the Milky Way from the radio \citep[e.g.][]{Beuther2016, Still2006}, to the sub-mm \citep[e.g.][]{Schuller2009, Aguirre2011, Rosolowsky2010}, the far-IR \citep[e.g.][]{Molinari2010}, the mid-IR \citep[e.g.][]{Benjamin2003}, and beyond.

Within our Galaxy, the inner few hundred parsecs have drawn considerable interest. This region, the Central Molecular Zone (CMZ), contains the largest reservoir of dense molecular gas and is distinct from the rest of the Galaxy in its extreme physical properties \citep[e.g.][]{Morris1996,Henshaw2023}. The CMZ is at a distance of 8.2 kpc \citep{Reid2019, Gravity18, Gravity19}. Gas from the Galaxy is funneled into the CMZ along the bar in `dust lanes' which then deposit the gas into the CMZ \citep[e.g.][]{Su2024, McClure-Griffiths2012, Hatchfield2020, Sormani2019}. Gas in the CMZ has been well-documented to be much more dense, hot, and turbulent than gas in the Galactic disk \citep[e.g.][]{Mills2013, Mills2018b, Ginsburg2016, Krieger2017, Kauffmann2017a, Federrath2016}, with stronger magnetic fields \citep[e.g.][]{Butterfield2024, Pillai2015}.

While theoretical models reproduce many features of the observed CMZ \citep[e.g.][]{Sormani2020, Tress2020, Hatchfield2020, Armillotta2019, Kruijssen2019,  Krumholz2017, Kruijssen2015, Krumholz2015}, many open questions remain about the morphology and kinematic properties of the gas and resulting star formation in the Galaxy's center. These issues, which are reviewed in detail in \citet{Henshaw2023}, include whether or not the CMZ has an episodic star formation cycle, the reason for a suppressed CMZ SFR despite its prodigious amount of dense gas, and the origin and nature of CMZ turbulence. Central to these questions is a measure of the dense gas clouds in the CMZ and their 3-D distribution. Since we lie in the disk plane, we only view the Galactic Center edge-on, and while gas kinematics \citep[e.g.][]{Henshaw2016b, Tsuboi1999, Sofue1995} can give some clue as to its 3-D structure, key questions remain about the general orbital shape of the gas and the  physical relationship between different parts of the CMZ.

In order to address these questions, observations at long wavelengths are essential in order to peer through the high extinction \citep[e.g.][]{Schodel2014} and robustly map out the densest CMZ gas.
Many continuum studies at (sub)mm wavelengths have been conducted of the CMZ \citep[e.g.~ATLASGAL, BGPS, and targeted surveys;][]{Schuller2009, Ginsburg2013, Bally2010, Tang2021a,Tang2021b, Ginsburg2020}, however from a single flux measurement it is not possible to unambiguously resolve the well-known degeneracy between temperature and column density. In this regard, the Herschel HiGAL survey \citep{Molinari2010, Molinari2011, Molinari2016} has a considerable advantage. HiGAL observed the Galactic plane in 5 filters from 70 to 500 \micron~making it possible to determine the continuum spectral energy distribution (SED) along every line of sight and thereby simultaneously fit for both the temperature and column density. We improve on previous analyses \citep[e.g.][]{Molinari2011, Tang2021a, Tang2021b} by performing a consistent fore/background subtraction and SED fit across the entire inner 40\deg~of the Galaxy.
 
This paper is the first in a series whose aim is to uncover the 3-D structure of our CMZ by cataloging key CMZ structures and assessing the near-far distance likelihood of each.
In this paper we describe the fore/background subtraction and SED fitting of Herschel HiGAL data towards the inner 40\deg~of the Galaxy and derive the temperature and column density distribution of the dense gas. In Paper II (Battersby et al., submitted) we segment these column density maps into their hierarchical substructures and place the CMZ in a global context. In Paper III (Walker et al., submitted) we perform a detailed kinematic analysis of cloud structures in the CMZ to identify contiguous structures in position-position-velocity space. The detailed kinematic analysis of Walker et al. (submitted) also compares molecular line tracer emission and absorption to help place these clouds on either the near- or far-side of the CMZ. Paper IV (Lipman et al., submitted) compares the Herschel dust emission for each cloud with its Spitzer 8\micron~and Herschel 70\micron~dust extinction to help constrain the 3-D position of clouds in the CMZ. 

This paper is organized as follows. In Section \ref{sec:data} we describe the data used in this work. In Section \ref{sec:methods} we describe the methods by which we subtract a cirrus fore/background and derive dust temperature and column density maps and their uncertainties. We also describe the public release of all data products in Section \ref{sec:release}. In Section \ref{sec:results} we present initial results of the SED fits, report on overall properties and trends in these data, and compare with previous work. In Section \ref{sec:analysis}, we measure the total dense gas mass, asymmetry, and extent of the CMZ, identify the warm dust ridge, and compare with previous work. We conclude in Section \ref{sec:conclusions}.

\section{Data}
\label{sec:data}

\begin{figure*}
\centering
\includegraphics[width=1\textwidth, trim={4mm 5mm 17mm 5mm}, clip]{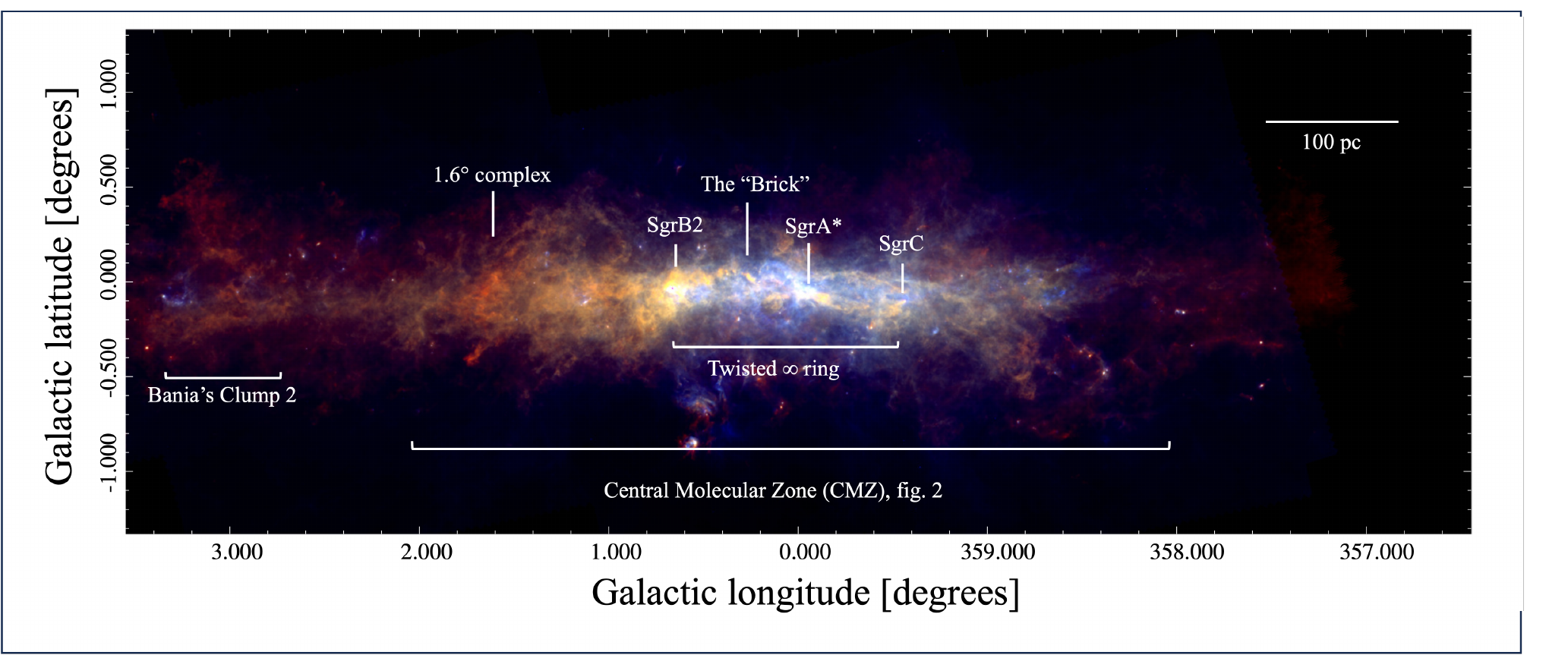}
\caption{A Herschel view of the inner 7\deg~of the Galaxy, Red: 350 \micron, Green: 160 \micron, Blue: 70 \micron. The main CMZ is clearly apparent by the bright concentration of emission within the inner 4\deg~longitude and 1\deg~latitude, but there are additional features, such as Bania's Clump 2 at $\ell$ $\sim$ 3.2\deg~at higher longitudes. The twisted inner ring \citep{Molinari2011a} is apparent in the inner 2\deg~longitude, and some flaring up to higher latitudes is seen outside this.}
\label{fig:herschelcmz}
\end{figure*}

The data used in this study were observed with Herschel as part of the Hi-GAL Survey \citep[Herschel Infrared Galactic Plane Survey;][]{Molinari2010, Molinari2016}, which covers the Galactic Plane at $|\ell|$ $\le$ 70\deg~ and $|b|$ $\le$ 1\deg.
The observing strategy and data reduction for Hi-GAL generally are described in \citet{Molinari2010} and \citet{Traficante2011} and for the CMZ specifically in \citet{Molinari2011a}. Briefly, the Spectral and Photometric Imaging Receiver \citep[SPIRE;][]{Griffin2010} and the Photodetector Array Camera and Spectrometer \citep[PACS;][]{Poglitsch2010} aboard Herschel observed the central region of the Galaxy in Parallel mode in five photometric bands centered at 70, 160, 250, 350, and 500 \micron. The corresponding beam sizes are about 6\arcsec, 12\arcsec, 18\arcsec, 25\arcsec, and 36\arcsec, respectively \citep{Molinari2016}. The zero level offsets in the images were derived by comparison with Planck/IRAS data \citep[as in][]{Bernard2010}, and are believed to be accurate within about 15\% \citep{Molinari2011a}. The Hi-GAL data products were released publicly at \href{http://vialactea.iaps.inaf.it}{http://vialactea.iaps.inaf.it}, as described in \citet{Molinari2016}. The Herschel view of our CMZ is shown in Figure \ref{fig:herschelcmz}.

\section{Methods}
\label{sec:methods}

In this Section we outline the methods for removing fore/background emission, deriving dust temperatures and column density maps and we describe uncertainties in these methods and provide details on the public data release. For all of our analyses we assume a distance to the CMZ of 8.2 kpc \citep{Reid2019, Gravity18, Gravity19}.

\subsection{Methods to Derive the Dust Temperature and Column Density Maps}
\label{sec:nh2_method}

\subsubsection{Overview of method}
\label{sec:overview}

The ISM is a mix of dust and gas in different phases and at different temperatures. In this work, we determine dust temperatures and column densities of the cold, molecular gas (T $\lessapprox 30$ K, traced by dust) component only by performing modified blackbody fits to Herschel data at the HiGAL-observed wavelengths. We direct the interested reader to the work of \citet{Barnes2017} who integrated Spitzer data with these Herschel data to measure both the warm and cold dust components within the inner CMZ. We also note that a number of publications introduce more sophisticated fitting approaches for determining the dust temperature and column density toward the CMZ using Herschel data, namely the PPMAP approach from \citet{Marsh2015, Marsh2016} as well as the Bayesian fitting approach from \citet{Tang2021a, Tang2021b} which also incorporates data from the LMT using AzTEC. We provide a global qualitative comparison with these works in Section \ref{sec:previous}. 

Our comparatively simple approach  is driven both by the availability of suitable datasets across the inner 40\deg~of the Galaxy as well as by the fact that the intrinsic uncertainties of this analysis (such as the dust opacity, assuming only one temperature component, see more in Section \ref{sec:uncertainties}) outweigh our measurement uncertainties. While it is possible to reduce uncertainties by attending to the detailed properties of a specific region, our approach to solve for the properties of the entire inner 40\deg~of the Galaxy negates this possibility on a first pass. Therefore, considering the high degree of intrinsic uncertainty to any fitting approach, we aim to apply a relatively simple fitting method over a very large area, and refer the reader to more detailed and complex approaches in targeted regions \citep[e.g.][]{Schmiedeke2016, Marsh2015, Etxaluze2011}.

At these wavelengths, the contribution of Galactic cirrus emission is significant and removal of this diffuse fore/background component is necessary.
We perform the cirrus subtraction and modified blackbody fitting using the automated methods described in \citet{Battersby2011} with one minor change to use an automated threshold for the number of iterations (detailed below). The methods are described briefly below, but we refer the reader to \citet{Battersby2011} for a more complete description of the method, including figures. The first step is to convolve and re-grid all of the images to a common resolution (36\arcsec), pixel-scale, and flux scale (MJy/sr).

These data products have been used already in \citet{Longmore2012}, \citet{Walker2015}, \citet{Henshaw2016a}, \citet{Henshaw2016b}, \citet{Mills2017}, \citet{Barnes2017}, \citet{Barnes2019}, \citet{Henshaw2020}, \citet{Battersby2020}, \citet{Hatchfield2020}, \citet{Hatchfield2024}.

\subsubsection{Source Masks and Cirrus Emission}
\label{sec:cirrus}
One of the first steps is to identify regions of significant emission associated with dense clumps and cores (we term these our `source masks') rather than Galactic cirrus.
The Galactic disk abounds with low density gas and dust between the stars \citep{Tielens2010, Draine2011, Klessen2016}. Far-IR radiation is emitted by the dust grains in this diffuse, high-Galactic latitude medium, known as the Milky Way cirrus. This cirrus emission was first identified by IRAS \citep{Hauser1984, Low1984}, and shortly thereafter, was seen to be strongly correlated with the atomic gas column density \citep{Boulanger1988} and heated by the general interstellar radiation field. This cirrus emission has been observed and analyzed in many studies since \citep[e.g.][]{Bianchi2017, PlanckCollaboration2014}. This emission is seen in Herschel wavebands \citep[e.g.][]{Martin2010} and in this work is excluded from the source masks and analyzed separately. For our analysis of the cirrus emission (Section \ref{sec:cirrus}) we include the 70 \micron~data point, however, for the analysis of structures within the source masks, this data point is excluded as it may be contaminated by the emission from very small dust grains \citep[e.g.][]{Desert1990, Compiegne2010}. Additionally, as described in detail in Paper IV (Lipman et al., submitted) 70 \micron~is seen in extinction toward a number of regions in the CMZ and would not be suitable as a measure of the dust emission.

The source masks and cirrus emission are identified through an iterative, automated routine that defines progressively more complete source masks with each iteration. This process is adapted from \citet{Battersby2011} and is described in full in that paper (Section 3.1, Figures 2-5). We summarize the overall procedure below, as well as a few minor updates. As in \citet{Battersby2011}, we identify regions of significant and contiguous emission (``source masks") at 500 \micron~where the contamination from cirrus emission decreases \citep{Gautier1992, Peretto2010a}. The final cirrus emission map is the remaining emission outside the final source masks smoothed to large scales.

As the first iteration, we smooth the Hi-GAL 500 \micron~image to 12\arcmin. After attempting a number of cirrus identification schemes in \citet{Battersby2011} we found that a 1-D Gaussian fit in Galactic latitude provided a reasonable fit to the cirrus emission over the Galactic Plane from $b=-1$\deg~to $+1$\deg. Therefore, we fit a 1-D Gaussian in Galactic latitude to each point in Galactic longitude of the smoothed image to construct a Gaussian fit image (at each Galactic longitude, the best Gaussian fit in latitude fills the latitude axis). We subtract the Gaussian fit image from the original 500 \micron~image to produce a difference image. This difference image is then used to estimate the noise ($\sigma_{\rm{rms}}$) in the following way. In this image, positive values contain signal as well as noise, while the negative values should be noise only. Therefore, we mirror the negative values about zero and fit a 1-D Gaussian to the flux distribution, which allows for an automated, approximate estimate of the noise ($\sigma_{\rm{rms}}$) in each difference image. We then apply a cutoff of 4.25$\sigma_{\rm{rms}}$\footnote{We tried cutoffs between 3 to 6 $\sigma_{\rm{rms}}$ in quarter-$\sigma_{\rm{rms}}$ intervals and 4.25$\sigma_{\rm{rms}}$ matched sources picked by eye best. The exact value is unimportant, the properties vary smoothly above and below this value} to the difference image. Pixels above this threshold define the first iteration source masks. 

In the next iteration, cirrus emission is determined by smoothing the 500 \micron~image, excluding the pixels inside the source masks, and each step of the process repeats. In \citet{Battersby2011}, we performed 16 iterations to reach the final source masks. However, the complexity and variation within the inner 40\deg~of the Galaxy for this work was better-suited to an automated approach that could adapt to more or fewer iterations as needed. We implemented an iterative cirrus emission determination until the source mask cutoff value converges. This convergence is defined as when the current mask cutoff and the two iterations prior have a standard deviation of less than 1 MJy/sr, the lowest RMS in the relevant Herschel 500 \micron~data \citep{Molinari2016}. We also implement a minimum of ten iterations, after which the cutoff varies very little as demonstrated in \citet{Battersby2011} Figure 5. We then determine the final cirrus emission maps by smoothing the Hi-GAL maps from 70-500\micron~to 12\arcmin~resolution, excluding the final source masks. 

These smoothed cirrus emission maps at each wavelength are then used to produce a fore/background column density and temperature map, using the same modified blackbody fit process described in the following section. The cirrus emission maps at each wavelength are subtracted from the original data maps to produce the source maps used for the modified blackbody fitting described in Section \ref{sec:bbfits}. The released data products show the cirrus-subtracted values inside the source masks and the smoothed cirrus emission outside the source masks. The label map denotes pixels inside the source masks with a 1 and pixels outside with a 0. 

\begin{figure*}
\includegraphics[angle=270, scale=0.75, trim={0cm 2cm 0cm 0cm},clip]{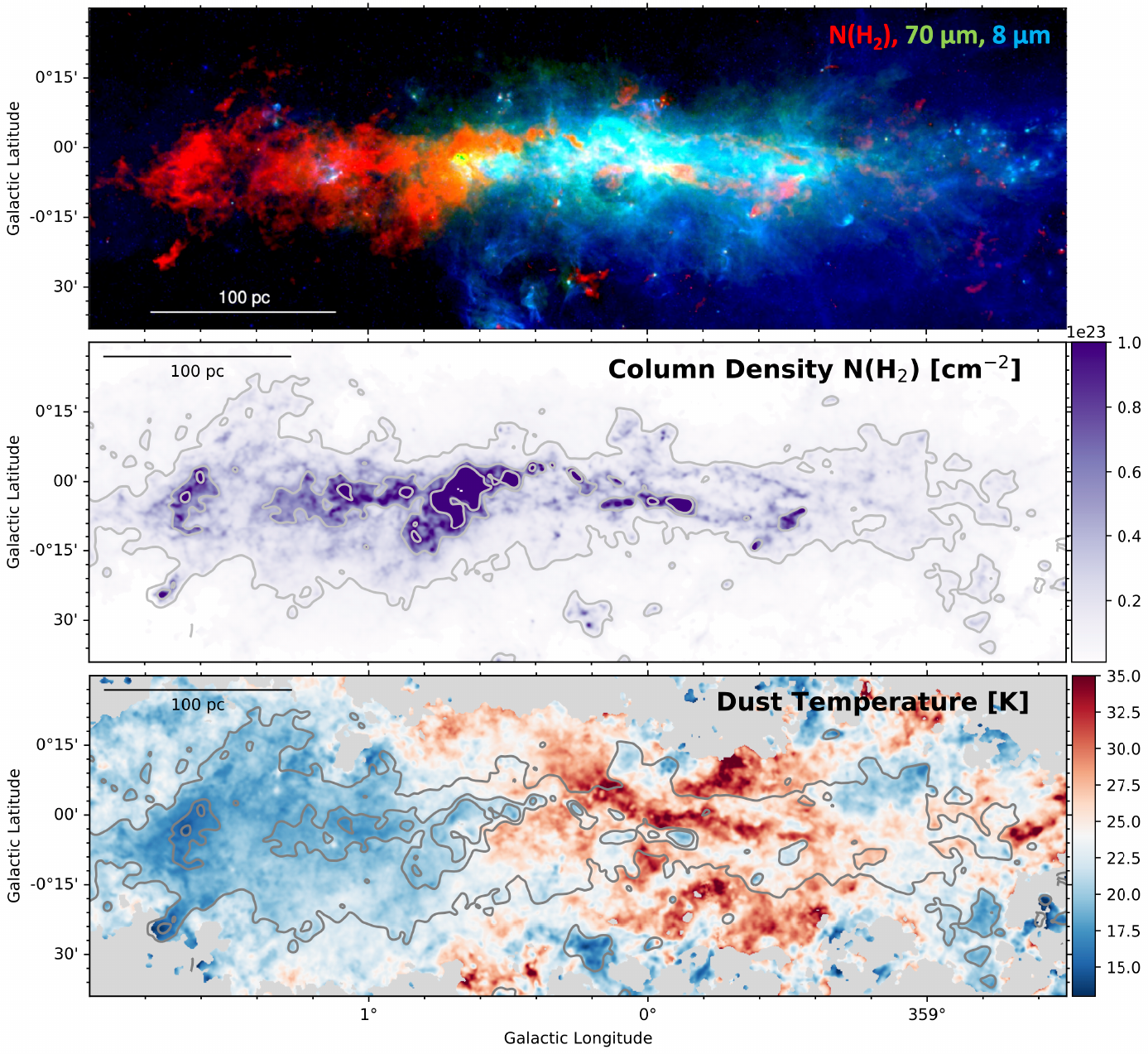}
\caption{The Central Molecular Zone (CMZ) contains an abundance of dense molecular gas that is highly asymmetric. This figure presents the CMZ as seen in a three-color compilation (top), in column density (middle), and dust temperature (bottom) from this work. \textit{Top:} Red is N(H$_2$) (this work), Green: 70 \micron~\citep[Hi-GAL;][]{Molinari2011a, Molinari2011b}, Blue: 8 \micron~\citep[GLIMPSE;][]{Benjamin2003}.  
\textit{Middle:}
Map of the column density derived in this work over the CMZ shows a high degree of asymmetry, with the majority of dense gas at positive longitudes (see \S \ref{sec:asymmetry}) and a majority of mid-IR emission is at negative longitudes. We also note the somewhat flatter scale height of dense gas in the inner degree that fans out at higher longitudes.
\textit{Bottom:}
Map of the dust temperature derived in this work over the CMZ show generally low dust temperatures of about 15-20 K, with a peak near the center of the Galaxy with an interesting dust ridge morphology discussed more in \S \ref{sec:ridge} and then decreasing towards positive longitudes. The temperature is generally lower at higher column densities (see Figure \ref{fig:nh2_temp_ridge}).
These maps are subject to a number of uncertainties and particular care should be taken in the interpretation of points near the edges of source masks (as described in \S \ref{sec:uncertainties}). Contours on the middle and bottom plots indicate N(H$_2$) = [1,5,10] $\times$ 10$^{22}$ cm$^{-2}$. The resolution of the column density and dust temperature maps are 36\arcsec~(or 1.4 pc at the assumed CMZ distance) as described in Section \ref{sec:overview}}.
\label{fig:3color_nh2_temp}
\end{figure*}

\subsubsection{Modified Blackbody Fits within the Source Masks}
\label{sec:bbfits}

We perform modified blackbody fits to each pixel within the source masks (after the cirrus emission has been subtracted). We include the Hi-GAL 160, 250, 350, and 500\micron~data in these fits. These data were convolved to the same resolution (36\arcsec), pixel scale, and flux scale (MJy/sr). A detailed discussion of uncertainties, both statistical and systematic, as well as assumptions in applying this method can be found in Section \ref{sec:uncertainties}. Implicitly assumed in this work is that all of our lines of sight (within the source regions) are dominated by H$_2$, which is reasonable given the high column densities in the CMZ \citep[e.g.][]{Mills2017}. 
We exclude the 70\micron~point as it may be contaminated by the emission from very small grains whose temperature distribution may differ substantially from that of the large grains probed by the other Herschel wavelengths \citep[e.g.][]{Desert1990, Compiegne2010}. 
We use the modified blackbody expression in the form
\begin{equation}
\label{eq:graybody}
S_{\nu} = \frac{2h\nu^{3}}{c^{2} (e^{\frac{h\nu}{kT}} - 1)}
(1 - e^{-\tau_{\nu}})
\end{equation}
where 
\begin{equation}
\label{eq:tau}
\tau_{\nu} = \rm \mu_{H_{2}} m_{H} \kappa_{\nu} \gamma \rm N(H_{2}) 
\end{equation}
The variables in equation 1 are: $S_\nu$ is the flux density as a function of frequency, $\nu$, in units of Watts per steradian per square meters per Hz. $h$ is Planck's constant, $c$ is the speed of light, $k$ is the Boltzmann constant, and $T$ is the temperature of the material. $\tau_\nu$ is defined by Equation 2 where
$\mu_{\rm H_{2}}$ is the mean molecular weight for which we adopt a 
value of $\mu_{\rm H_{2}}$ = 2.8 \citep{Kauffmann2008}, m$_{\rm H}$ is the mass of a hydrogen atom, N(H$_{2}$) is the H$_{2}$ column density,  $\kappa_{\nu}$ is the dust opacity, and $\gamma$ is the gas-to-dust ratio. For the gas to dust ratio, we adopt a standard value of 100. While this gas to dust ratio is commonly used \citep[e.g.][]{Barnes2017, Battersby2010, Battersby2011}, it is a largely unknown quantity, without even clear uncertainties on the value \citep[e.g.][]{Lv2017}, see Section \ref{sec:uncertainties} for more details). As in \citet{Battersby2011}, we derive a dust opacity as a continuous function of frequency by fitting a power-law to the tabulated \citet{Ossenkopf1994} dust opacities over the relevant range of frequencies. The \citet{Ossenkopf1994} opacities assume an MRN (Mathis, Rumpl, and Nordsieck) grain size distribution \citep{mrn77}. Specifically, we use the version of the Ossenkopf \& Henning model that assumes that the grains have thin ice mantles and have undergone 10$^5$ years of coagulation at a density of 10$^6$ cm$^{-3}$, which is a reasonable and common assumption for these cold, dense clumps.
This yields an expression for the dust opacity of 
\begin{math} \kappa_{\nu} = \kappa_{0} (\nu/\nu_{0}) ^{\beta} \end{math}where $\kappa_{0}$ is 4.0 cm$^2$ g$^{-1}$ for $\nu_{0} = 505$~GHz with a fixed $\beta$ equal to 1.75, between the standard values of 1.5-2.
For this analysis we opted for a fixed $\beta$ considering the limited data available and uncertainties (see Section \ref{sec:uncertainties}), but we direct the reader to \citet{Tang2021a} for work suggesting a dust spectral index of between $2 - 2.4$.

We used the package {\sc montage}\footnote{\href{http://montage.ipac.caltech.edu/}{http://montage.ipac.caltech.edu/}} to stitch and re-grid together images from the original Hi-GAL released 2\deg$\!\times$2\deg~tiles. 
For most of the analysis, we use a mosaicked image containing the inner 7 degrees (in longitude, the latitude extent is the Hi-GAL coverage from $b=-1$\deg~to +1\deg) of the Galaxy, from $\ell$ = +3.5\deg~to -3.5\deg. This means that the automated cirrus fore/background identification and subtraction was performed over this entire region. We also mosaicked together the inner 40 degrees in longitude of the Galaxy from $\ell$ = +20\deg~to -20\deg, and performed cirrus fore/background identification and subtraction on this region. This complete inner 40\deg~map is used in Figure \ref{fig:inner40}, while the inner 7\deg~map is used in the rest of the analysis. The resulting source-identified regions and column densities unsurprisingly show slight variations between the two maps in the overlapping region, due to slightly different automated fore/background identification, however, these differences are well within the quoted uncertainties (see Section \ref{sec:uncertainties}).

\subsubsection{Uncertainties in the Temperatures and Column Densities}
\label{sec:uncertainties}


The formal statistical errors on the derived dust temperatures and column densities are strong functions of the parameter values, with a median temperature error of 11\% and a median column density error of 25\%. These are the errors in the derived properties assuming 20\% calibration uncertainty in the Hi-GAL fluxes \citep{Molinari2010}. However, the intrinsic and systematic errors are much higher. The definition and subtraction of the cirrus emission is a unique source of error for this analysis of Herschel data, since its structure and magnitude are not known a priori. Other major sources of uncertainty are the  assumptions of the dust opacity and the gas to dust ratio. We have very limited data on the validity of the commonly used gas to dust ratio of 100 \citep[e.g.][]{Lv2017}. Due to the different temperatures, densities, and other physical properties in the Galactic Center \citep[e.g.][]{Henshaw2023}, it is quite plausible that the gas to dust ratio may be different, which would have a large impact on the inferred quantities. The gas to dust ratio may be lower if there is a higher metallicity at the Galactic Center, which would affect our mass and density estimates linearly. Based on the Monte-Carlo error simulations of \citet{Battersby2010} using similar analyses, we expect that our measured column densities have a systematic uncertainty of about a factor of two. However, the \citet{Battersby2010} Monte Carlo simulations did not include all possible sources of errors, including the gas to dust ratio. So, the true systematic uncertainty may be even higher. The relative uncertainty between regions in our maps, however, is smaller, since many of these systematic uncertainties (such as the choice of gas to dust ratio) apply equally to all regions.

Additionally, the ability of the 160 - 500 \micron~data to accurately capture dust temperatures above 30 K is quite limited, as these data are really only sensitive to the cold dust component with these fits. Therefore, in warmer temperature regions, we would expect to have higher uncertainties, and if the temperature is under-estimated in our fits, then the column density would be over-estimated. We also note that since the `sources' we identify flow gradually into the cirrus emission, pixels near the edges of the source masks are highly uncertain. 

Finally, these measurements are also uncertain due to the inherent assumption that a single-temperature modified blackbody is a sufficient approximation to the physical environment. There certainly exist temperature and column density fluctuations on scales smaller than our beam, and along the line-of-sight. These measurements, therefore, represent the average properties of the regions along the line-of-sight and within the beam.

\subsubsection{Data Release}
\label{sec:release}

We release all the analysis products from this work (including fore/background-subtracted column density and dust temperature maps over the inner 40\deg~of the Galaxy) 
in the 3-D CMZ Harvard Dataverse: \href{https://dataverse.harvard.edu/dataverse/3D_CMZ}{https://dataverse.harvard.edu/dataverse/3D\textunderscore CMZ}. The dataset for this paper is found in the Papers I and II repository: \href{https://doi.org/10.7910/DVN/7DOJG5}{https://doi.org/10.7910/DVN/7DOJG5} \citep{Battersby_3DCMZII_data}. The Harvard Dataverse is an online data repository that enables long-term data preservation and sharing. The details of the data products available are described in Appendix \ref{sec:appendix_data} as well as in the online meta-data. Project updates can also be found on the 3-D CMZ website:  \href{https://centralmolecularzone.github.io/3D_CMZ/}{https://centralmolecularzone.github.io/3D\textunderscore CMZ/}.

\section{Results}
\label{sec:results}

We display the column density and dust temperature maps derived in this work in the inner $\sim$3.5\deg~longitude of the Galaxy in Figure \ref{fig:3color_nh2_temp} and discuss these results in Sections \ref{sec:nh2} and \ref{sec:temp}. We compare dust temperatures across the CMZ in Figure \ref{fig:temp_pdf} and Section \ref{sec:temp}. We describe the results across the full inner 40\deg~longitude in Section \ref{sec:inner40} and in Figure \ref{fig:inner40}. We describe key insights from these results on the CMZ asymmetry,  extent, and total dense gas mass in Sections \ref{sec:asymmetry}, \ref{sec:extent} and \ref{sec:mass}. We identify a ridge of warm dust in Section \ref{sec:ridge} and display it in Figure \ref{fig:meerkat}. We also present a 2-D histogram of the column density and dust temperature points for the CMZ in Figure \ref{fig:nh2_temp_ridge} with the warm dust ridge highlighted.
In Section \ref{sec:previous} we compare with previous work.

\subsection{Column Density in the CMZ}
\label{sec:nh2}
The column density map over the inner 3.5\deg~longitude of the Galaxy is shown in Figure \ref{fig:3color_nh2_temp}.
The peak column density in the map is toward the SgrB2 complex with a value of about 2 $\times$ 10$^{24}$  cm$^{-2}$. However, the peak density is observed towards several pixels which are saturated in the Herschel data towards Sgr B2 North and Main {\footnote{Sgr B2 North and Main refer to the densest portions of the Sgr B2 molecular cloud, marked in Figure \ref{fig:herschelcmz}, and have coordinates of approximately $\alpha_{J2000} = 17h47m19.88s$, $\delta_{J2000} = -28^{\circ}22^{\prime}18.4^{\prime\prime}$ and $\alpha_{J2000} = 17h47m20.35s$, $\delta_{J2000} = -28^{\circ}23^{\prime}3.0^{\prime\prime}$, respectively, please see \citet{Schmiedeke2016} for more details.}, indicating a saturated pixels towards very bright emission. Missing this bright emission (likely associated with very high column density gas) and beam-smearing of the relatively coarse Herschel data both act to lower our measurement and the true column density in this region is certainly higher \citep[as demonstrated with higher resolution data, e.g.][]{Hatchfield2020, Ginsburg2018, Schmiedeke2016}. Typically ``clouds" in the CMZ have edges at about N(H$_2$) $\sim$10$^{23}$ cm$^{-2}$ \citep[this threshold was used to define the boundary of CMZ clouds in the CMZoom survey; see][]{Battersby2020} with peaks of a few times 10$^{23}$ cm$^{-2}$ in their centers. 

These column densities are sufficient to absorb a substantial fraction of the soft  (E $<$ 4.5 keV) X-ray emission in the Galactic center, as can be seen in deep \textit{XMM-Newton} images \cite{Ponti15}, where clouds like Sgr B2, the Brick, and Sgr C appear in silhouette. The CMZ is unique in our galaxy for having an extremely large molecular gas fraction \citep[$N_{H2}/N_H >90$\%;][]{Sofue22}. As a result, the Compton thickness of clouds in this region is most correctly calculated relative to the column density of molecular hydrogen. Taking the cross section of H$_2$ to be 2.8 larger than H at high energies \citep{Yan98}, CMZ clouds would become Compton thick for $N$(H$_2$) $>$5.4 $\times 10^{23}$ cm$^{-2}$. While only the Sgr B2 cloud exceeds this limit in our Galaxy's center at these resolutions, it is likely higher on smaller spatial scales, and a number of other regions are within a factor of two (our estimated uncertainty) of this value. Other galaxies may have even denser central regions \citep[e.g.][]{Stuber2023}. If a substantial fraction of the gas exceeds these limits, and if the viewing angle towards a central supermassive black hole is obscured by this dense gas, then gas in the central $R\sim200$ pc of galaxies could play in obscuring extragalactic AGN.

\begin{figure*}
\centering
\subfigure{
\includegraphics[width=0.9\textwidth]{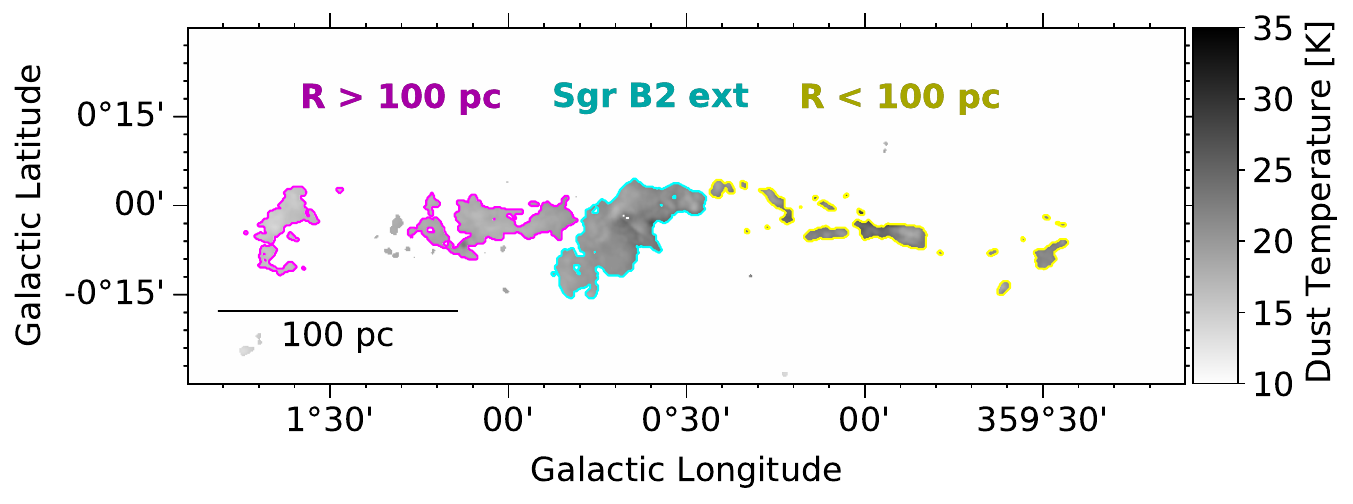}}
\subfigure{
\includegraphics[width=0.8\textwidth]{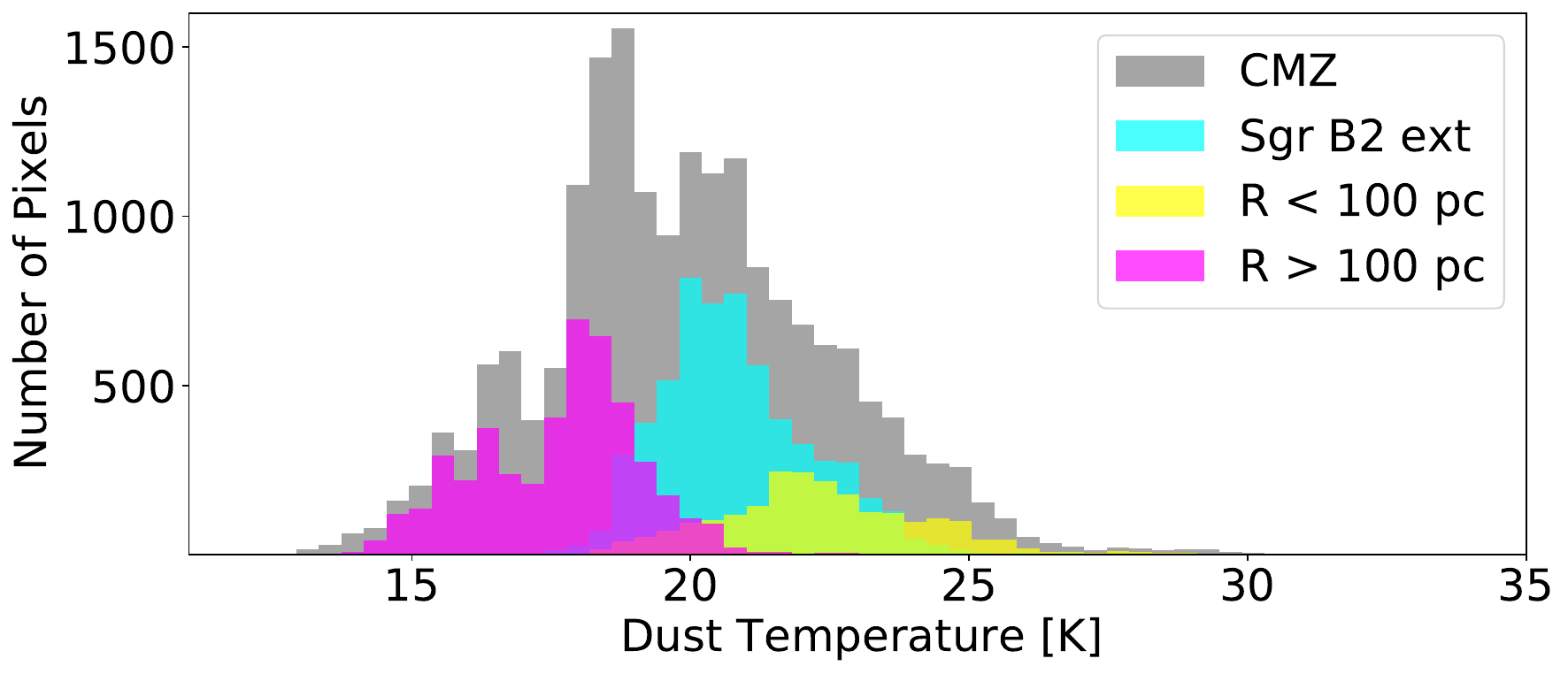}}\\
\caption{We find that dust temperature in the CMZ shows relatively small variation (from about 12 - 35 K) but is the highest in the inner CMZ, followed by what we call the Sgr B2 extended region, and lowest at higher Galactic longitudes. 
These rough delineations are defined by natural separations in the column density map as described in Section \ref{sec:temp} and do not correspond perfectly to Galactocentric radius or the SgrB2 region, but are meant to be characteristic approximations for describing general trends.
The \textit{top panel} presents a map of our derived CMZ dust temperatures (for pixels where N(H$_2$) $\ge$ $6 \times 10^{22}$ cm$^{-2}$) and the delineation of three CMZ regions (inner CMZ in yellow, SgrB2 extended in cyan, and outer CMZ in magenta). The \textit{bottom panel} shows the histogram of pixels in the above map as a function of temperature, showing again the overall CMZ in gray, inner CMZ in yellow, Sgr B2 extended in cyan, and the outer CMZ in magenta.}
\label{fig:temp_pdf}
\end{figure*}

\subsection{Dust Temperature in the CMZ}
\label{sec:temp}

In this work, we report on the dust temperature in the CMZ. In molecular clouds, it is generally thought that at high densities (n $\ge 10^{4.5}$ cm$^{-3}$), molecular gas and dust particles are sufficiently coupled, such that their temperatures should be well-matched \citep[e.g.][]{Goldsmith2001, Young2004}. However, it is well-documented that in the CMZ the measured gas temperatures are significantly higher than the dust temperatures \citep[e.g.][]{Marsh2016, Tang2021a, Henshaw2023} despite their high densities. This was reproduced in simulations by \citet{Clark2013} and is also seen in other galaxy centers \citep[e.g.][]{Mangum2013}. Therefore, we do not expect that our dust temperatures necessarily correspond to the molecular gas temperatures in this region; however, they serve as an independent tracer of temperature variation trends and describe the thermal state of the dust component in this region.

The dust temperature map in the bottom panel of Figure \ref{fig:3color_nh2_temp} shows the range in dust temperatures derived in this work\footnote{There exist spurious pixels near the source edges with temperatures ranging from 5 to 45 K, but the majority (99.5\%) are within the stated range of 12 to 35 K.}, from about 12 K to about 35 K with a median dust temperature of 21 K and a standard deviation of 4 K. The highest dust temperatures are in the very central regions of the CMZ, but avoid the dense cloud regions. 

Figure \ref{fig:temp_pdf} shows both a map and histogram of dust temperature across the CMZ. For this analysis, we divide the CMZ into three approximate regions based on natural separations in the column density map, in this case at N(H$_2$) = $6 \times 10^{22}$ cm$^{-2}$. The three resulting regions are approximately R $<$ 100 pc (0.4\deg~ $\gtrsim$ $\ell$ $\gtrsim$ -0.6\deg), SgrB2 (0.8\deg~ $\gtrsim$ $\ell$ $\gtrsim$ 0.4\deg), and R$>$100 pc (1.8\deg~ $\gtrsim$ $\ell$ $\gtrsim$ 0.8\deg), but the exact regions are shown as contours in Figure \ref{fig:temp_pdf}. The top panel shows the dust temperature map in grayscale (for pixels where N(H$_2$) $\ge$ $6 \times 10^{22}$ cm$^{-2}$) with the three regions shown as colored contour edges. The bottom panel shows the histogram of the temperature distribution for the CMZ overall (gray), the inner 100 pc (yellow), the SgrB2 region (cyan), and the outer 100 pc (magenta). We see that the highest dust temperatures are found in the inner 100 pc, followed by the SgrB2 region, with the lowest dust temperatures in the outer 100 pc region.

As can be seen in the bottom panel of Figure \ref{fig:3color_nh2_temp}, there is a ridge-like structure of increased dust temperature directly above SgrA* and roughly parallel to the Galactic plane, sloping to higher latitudes at about $\ell \sim$ 0.1\deg, and patches of high dust temperature directly above and below the Galactic Center, outside the dense cloud complexes. We discuss this region, and its potential association with Galactic outflows or massive stars in Section \ref{sec:ridge}. 

The bottom panel of Figure \ref{fig:inner40} displays the dust temperature for pixels with N(H$_2 > 10^{22}$ cm$^{-2}$ as a function of Galactic longitude, averaged along the latitude axis, with some negative longitude slices missing high enough column density. The dust temperature peaks at a Galactic longitude between $\ell = -0.5$\deg~to 0.5\deg, is generally higher at more negative Galactic longitudes and lower towards higher Galactic longitudes, with the minimum continuous region of low dust temperature in the 1.6\deg~cloud complex. The bottom two panels of Figure \ref{fig:3color_nh2_temp} display the same N(H$_2$) contours. The second contour level at a column density of  5 $\times 10^{22}$ cm$^{-2}$ corresponds to uniformly low dust temperatures (T $<$ 25 K) and defines the key cloud structures within the CMZ. 

It is well-documented that typical modified blackbody fitting routines show an anti-correlation between high dust temperature and high column density \citep[e.g.][]{Battersby2011, Tang2021a}, which we also see here. We plot the column density vs. dust temperature in Figure \ref{fig:nh2_temp_ridge}.

\begin{figure*}
\centering
\includegraphics[trim={0 4mm 0 0}, clip, width=1\textwidth]{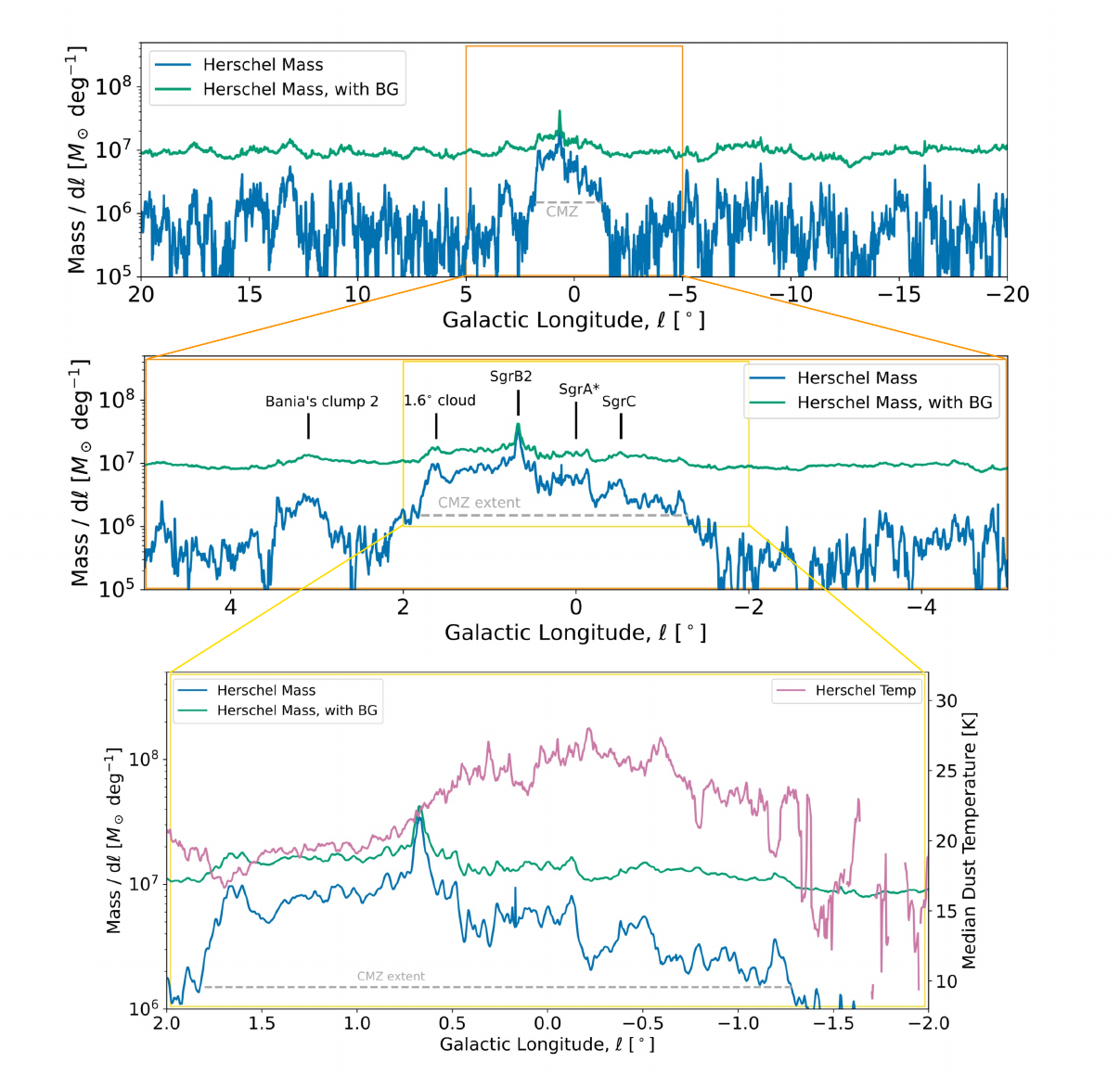} 
\label{fig:inner40}
\caption{The CMZ clearly stands out as the large, contiguous region of high column density gas in the inner 3\deg~of the Galaxy. We estimate a longitude extent of the CMZ of approximately 1.8\deg~$<\ell<$-1.3\deg~(shown as the dashed gray line, see Section \ref{sec:extent}). This figure displays the mass per unit Galactic longitude (summed over the observed latitude range of b=$\pm 1$\deg) derived in this work for the inner 40\deg~(\textit{top panel}), 10\deg~(\textit{middle panel}), and 3\deg~(\textit{bottom panel}) of the Galaxy. The dark blue line shows the mass derived using the fore/background subtracted column density maps and represents our best estimate of the true dense gas mass. The lighter green line has no fore/background subtraction (labeled Herschel mass, with BG) performed and therefore includes all the diffuse cirrus emission along the line-of-sight. The bottom panel also displays the median Herschel dust temperature at each longitude slice (for pixels with a column density of at least N(H$_2$) = 10$^{22}$ cm$^{-2}$), showing a modest average increase from the lowest dust temperatures at higher Galactic longitudes, to more negative Galactic longitudes. }
\end{figure*}

\subsection{The Inner 40\deg~of the Galaxy}
\label{sec:inner40}

As part of this work, we stitched together (using {\sc montage} as described in Section \ref{sec:nh2_method}) and created column density and temperature maps in a systematic way for the entire inner 40\deg~of the Galactic Plane using the Hi-GAL dataset. These are publicly released as described in Section \ref{sec:release}. The automated fore/background separation scheme was applied uniformly across this region and, through qualitative comparison with the ATLASGAL \citep{Csengeri2014} and BGPS \citep{Ginsburg2013} catalogs, appears to have been successful in picking out the highest column density structures. 

Perhaps the most striking aspect of the inner 40\deg~mosaics is how clearly the CMZ stands out as by far the largest contiguous region of high density gas in the entire inner Galaxy (discussed more in Section \ref{sec:extent}). This is most clearly seen in Figure \ref{fig:inner40}, which displays the differential mass per each longitude slice (summed over the observed latitude range $b = \pm 1$\deg~latitude) for the inner 40\deg~(top panel), inner 10\deg~(middle panel), and inner 4\deg~(bottom panel). The dark blue line is the calculated mass per longitude with our fore/background subtraction and most clearly demonstrates the large, contiguous high density region that is the CMZ. The lighter green line is with no fore/background subtraction, which shows a relatively flat mass profile with only relatively small variations. Even in this entire inner 40\deg~view, SgrB2 clearly stands out as the densest region of the inner Galaxy. 


\section{Analysis}
\label{sec:analysis}

\subsection{Insights on the 3-D CMZ from Dust Continuum: a Large Asymmetry}
\label{sec:asymmetry}

The CMZ is highly asymmetric in longitude and latitude, with the majority of dense gas at $\ell$ $>$ 0\deg and at $b <$ 0\deg. The latitude asymmetry is only partly explained by the Sun's position above the Galactic plane \citep[e.g.][]{Goodman2014, Battersby2011, Rosolowsky2010}, and instead may be caused by the tilt of the inner Galaxy with respect to the Milky Way's disk midplane \citep[see for example Figure 1 in][]{Sormani2019}.
We investigate the longitude asymmetry in Table \ref{tab:masses} across different CMZ regions, finding that about 70-75\% of dense gas (with the fore/background subtracted) in the CMZ is at positive longitudes. This value drops to as low as 50-60\% in the images including fore/background, highlighting that this asymmetry is restricted to the densest gas of the CMZ.

In the inner 7\deg~Herschel view of the CMZ in Figure \ref{fig:herschelcmz}, the immense brightness of the CMZ (from about 2\deg$<\ell<359$\deg\ and 0.5\deg$<b<-0.5$\deg) compared with the rest of the inner Galaxy is highlighted. The asymmetry of the CMZ is most prominent in the derived column density map seen in Figure \ref{fig:3color_nh2_temp} as well as in the differential mass plot in Figure \ref{fig:inner40}. The top panel of Figure \ref{fig:3color_nh2_temp} shows an overview of the CMZ, including Herschel-derived column densities (in red), Herschel 70 \micron~emission (green), GLIMPSE 8 \micron~emission (blue), and demonstrates intriguing color variation. The inner 100 pc contains diffuse, widespread mid-IR emission (blue and green) as well as dense gas (red). The dense core of SgrB2 shows bright emission at all wavelengths, containing dense, actively star-forming gas. However, outside this central region, the 1.6\deg~ clouds at higher longitudes show nearly pure red (dense gas) emission while toward more negative longitudes the mid-IR emission (blue and green) dominates and extends above and below the plane. 
Much of the $\ell$ $<$ 0\deg~ mid-IR emission is diffuse, but there also exist numerous compact sources, whose nature is debated \citep[e.g.][]{Yusef2009, Koepferl2015}. Some of these sources are known compact HII regions in the Sgr E region \citep{Anderson2020} and show complex morphologies that may indicate gas shearing as it enters the CMZ \citep{Wallace2022, Gramze2023}. The inner 100 pc `streams' \citep{kdl2015, Molinari2011a} can be seen as the infinity-like shape in the dense gas (red) in Figure \ref{fig:3color_nh2_temp}, and is highlighted in the \citet{Molinari2011a} overview of the CMZ seen with Herschel.

The asymmetry of the CMZ was first identified more than 30 years ago \citep{Bally1988} and has been noted many times since \citep[e.g.][]{Bally2010, Longmore2017, Henshaw2016a}. While its origin has been a topic of debate for some time, \citet{Sormani2018a} present a simple explanation for this asymmetry based on 3-D hydrodynamic simulations. They posit that the key factor is the unsteady flow of gas in a barred potential. Despite having a point-symmetric starting point, the hydrodynamical simulations of \citet{Sormani2018a} develop asymmetries due to unsteady gas flow which originates from a combination of the so-called wiggle instability \citep{Mandowara2022}, the thermal instability, and bombardment of the CMZ from the bar shocks. 
Since an asymmetry such as that observed in our CMZ (in both magnitude and position) is common at many timesteps in their simulations, they suggest that the present-day asymmetry is a transient feature that repeats often during the lifetime of the Milky Way on timescales of tens of Myr. Other mechanisms, such as stellar feedback, self-gravity, a live stellar potential, or pre-existing large-scale asymmetries could also contribute to create further asymmetries in the gas distribution, and their contribution is not well constrained. See Section 4.3.1 of \citet{Henshaw2023} for more discussion on this topic.

\subsection{Insights on the 3-D CMZ from Dust Continuum: Extent}
\label{sec:extent}

In both Figures \ref{fig:herschelcmz} and \ref{fig:inner40}, the CMZ clearly stands out as being denser and brighter than the rest of the inner Galaxy. It is the largest contiguous region of high column density gas within the inner 40\deg~of the Milky Way, and therefore is the largest contiguous region of high column density gas in the entire Galaxy. 

Using the data shown in Figure \ref{fig:inner40}, we estimate the average mass per longitude for the inner Galaxy to be about 5 $\times 10^5$ \Msun~deg$^{-1}$. We then identify the region that is 3$\times$ this value over a contiguous longitude range. This corresponds to a longitude range of about 1.8\deg $> \ell >$ -1.3\deg. Based on the contiguity of dense gas and separation from the rest of the Galaxy as a significant overdensity, we conclude that the longitude extent of the CMZ is 1.8\deg $> \ell >$ -1.3\deg. Visual inspection indicates that the dense gas spans about $b=\pm 0.75$\deg in latitude space.

Many previous works consider the CMZ to be only the inner 100 pc `streams' \citep[e.g.][]{Molinari2011,Kruijssen2015, Ginsburg2016, Krieger2017} of gas that extends from about 1\deg $<\ell<-0.7$\deg. However, \citet{Henshaw2023} and references within define this approximate region (2.0\deg $> \ell >$ -1.3\deg) as the \{$\ell, v$\} parallelogram which likely includes both the CMZ as well as its interaction points with the dust lanes. They therefore define the CMZ as the `mass-dominant component' over 1.7\deg $> \ell >$ -1.0\deg, which includes the 1.3\deg~complex. Our measurements of the dense gas over the inner 40\deg~of the Galaxy and observation of a clear, contiguous, and significant over-density of gas from about 1.8\deg $> \ell >$ -1.3\deg~supports the idea that the CMZ is wider than previously thought and connects with the \{$\ell, v$\} parallelogram.

\begin{table*}
\centering
\caption{Total mass of dense gas in the CMZ over different areas, calculated using Herschel data in this work. The main values are thought to be more reliable using the dense gas mass, however, the numbers in parentheses do not include fore/background subtraction and are contaminated by cirrus emission, are included for reference as an upper limit of the gas detected by Herschel. The last two columns are the percentage of the total mass that is at positive ($\ell > 0$\deg) vs negative ($\ell < 0$\deg) longitudes, and are only calculated for symmetric CMZ regions.}
\noindent\begin{tabular}{@{}lcccc}
\hline
CMZ Region & Physical Scale [pc]\footnote{Assuming a Galactic Center distance of 8.1 kpc \citep{Reid2019, Gravity19}} & Total Mass [\Msun] & Positive \% & Negative \% \\ 
\multicolumn{4}{l}{\textit{Main values are fore/background subtracted, ones in (parentheses) include cirrus emission}} \\
\hline
Inner 7\deg $\times$ 2\deg\footnote{All `inner' values are centered on [$\ell$,b]=[0\deg,0\deg]} & 1000 $\times$ 280 & 2.8 $\times 10^7$ (7.8 $\times 10^7$) & 74\% (58\%) & 26\% (42\%) \\
Inner 4\deg $\times$ 2\deg & 570 $\times$ 280 & 2.4 $\times 10^7$ (5.0 $\times 10^7$) & 74\% (60\%) & 26\% (40\%) \\
Inner 2\deg $\times$ 2\deg & 280 $\times$ 280 & 1.5 $\times 10^7$ (2.8 $\times 10^7$) & 69\% (58\%) & 31\% (42\%) \\
Inner 1.05\deg $\times$ 1.05\deg\footnote{To match the region from \citet{Ferriere2007}, who found a total mass of about 1.3 $\times 10^7$ \Msun} & 150 $\times$ 150 & 0.6 $\times 10^7$ (1.0 $\times 10^7$) & 56\% (52\%) & 44\% (48\%) \\
$\ell$ = -1.05\deg~to 2.25\deg, b = $\pm$ 0.75\deg\footnote{To match the region from \citet{Dahmen1998}, who found a total mass of $\sim 1.2-6.4 \times 10^7$ \Msun} & 470 $\times$ 210 & 2.3 $\times 10^7$ (4.1 $\times 10^7$) & -- & -- \\
$\ell$ = -1.0\deg~to 1.7\deg, b = $\pm$ 0.5\deg\footnote{To match the \citet{Henshaw2023} definition of the CMZ region} & 380 $\times$ 140 & 2.0 $\times 10^7$ (3.1 $\times 10^7$) & -- & -- \\
$\ell$ = -1.3\deg~to 1.8\deg, b = $\pm$ 0.75\deg\footnote{Extent measured from this work} & 440 $\times$ 210 & 2.2 $\times 10^7$ (3.9 $\times 10^7$) & -- & -- \\
\hline
\end{tabular}
\label{tab:masses}
\end{table*}

\subsection{The Total Dense Gas Mass of the CMZ}
\label{sec:mass}

The column density map of the CMZ can be integrated to estimate the total dense molecular gas mass of the CMZ. As described in Section \ref{sec:methods}, we use the fore/background subtracted maps for our main analysis, but include constraints from the maps with the fore/background included in Table \ref{tab:masses} as well.
We consider any gas with a measured column density of N(H$_2$) $> 10^{22}$ cm$^{-2}$ to be dense. In the translation of high column density dust seen with Herschel to the inference of dense, molecular gas are several assumptions: 1) high column density gas corresponds to high volume density gas and 2) dust seen with Herschel traces dense, molecular gas. These assumptions are supported by the many studies \citep[e.g.][]{Mills2018, Krieger2017, Ginsburg2016} which find abundant dense, molecular gas in the CMZ, with intensity peaks that largely correspond with the Herschel data \citep{Mills2017}. We estimate that most of the gas traced with Herschel observations is likely $n > 10^4 \: \mathrm{cm}^{-3}$ and we refer to this as `dense, molecular gas' throughout.

In order to estimate the total dense gas mass of the CMZ, we integrate the column density map.
We perform this integration over several different areas of the CMZ and present the results in Table \ref{tab:masses}. Over all the different areas, we find total dense gas mass values ranging from 0.6-2.8 $\times 10^7$ \Msun~in the nominal fore/background-subtracted maps, or 1.0 - 7.8 $\times 10^7$ \Msun~in the images including fore/background emission, likely contaminated by cirrus emission. From the carefully-defined CMZ area from the review by \citet{Henshaw2023} of $\ell = -1.0$\deg~to 1.7\deg, b = $\pm$ 0.5\deg, we find a total dense gas mass of the CMZ of 2.0 $\times 10^7$ \Msun~(3.1 $\times 10^7$ \Msun~without fore/background subtraction). Using the extent defined in this work of $\ell = -1.3$\deg~to 1.8\deg, b = $\pm$ 0.75\deg, we find a total dense gas mass of the CMZ of  2.2 $\times 10^7$ \Msun~(3.9 $\times 10^7$ \Msun~without fore/background subtraction), which we adopt as our standard value. 

The total dense gas mass values in Table \ref{tab:masses} are in line with the typically quoted values of about 2-6 $\times 10^7$ \Msun~ \citep[e.g.][]{Henshaw2023, Molinari2011a, Dahmen1998, Ferriere2007, Morris1996}. Table \ref{tab:masses} includes two regions for specific comparison with previous works. We first compare with \citet{Ferriere2007} who calculated a total molecular gas mass using $^{13}$CO in the inner 150 pc of the CMZ of about 1.3 $\times 10^7$ \Msun, updated from \citet{Sofue1995} with a new $X_{\rm CO}$ factor, compared with our value of 0.6 $\times 10^7$ \Msun (1.0 $\times 10^7$ \Msun~ including fore/background). We also compare with \citet{Dahmen1998} who calculate a CMZ mass over a region of $\ell = -1.05$\deg~to 2.25\deg, b = $\pm$ 0.75\deg~ and find a value of 1.2 $\times 10^7$ \Msun~ based on C$^{18}$O (with an upper limit of 6.4 $\times 10^7$ \Msun). Our mass in this same region is 2.3 $\times 10^7$ \Msun~ (4.1 $\times 10^7$ \Msun~ including fore/background). 

Astrophysical mass measurements are notoriously challenging and have errors that are largely dominated by systematic effects. These errors include the uncertainty in the dust opacity, the $X_{\rm CO}$ conversion factor, the separation of fore/background, and are estimated to have an overall systematic uncertainty of about a factor of two \citep{Battersby2010}. In this work, we present masses derived using far-IR dust continuum. These masses present a unique complement to the majority of mass estimates in the literature, which are based on CO (and isotopologue) emission. While the CO estimates have the clear benefit of being able to more carefully select regions by incorporating velocity information, it is also complicated by an uncertain $X_{\rm  CO}$ factor and other excitation uncertainties \citep[e.g.][]{Sandstrom2013}. The far-IR dust continuum provides a relatively simple counterpoint that has orthogonal benefits (relatively simple excitation process, optically thin emission) and drawbacks (limited to plane-of-sky information, dust opacity and dust-to-gas ratio uncertainties). Overall, we find an agreement within a factor of two between previous works using CO and dust continuum emission in this work. This agreement lends confidence to both approaches, and suggests that the total dense mass of our CMZ is well-established to be about $2\substack{+2 \\ -1} \times 10^7$ \Msun~ \citep[our value with a factor of two uncertainty, using the CMZ region defined in this work, and in agreement with the value derived in the region from][]{Henshaw2023}.

\begin{figure*}
\includegraphics[width=1\textwidth]{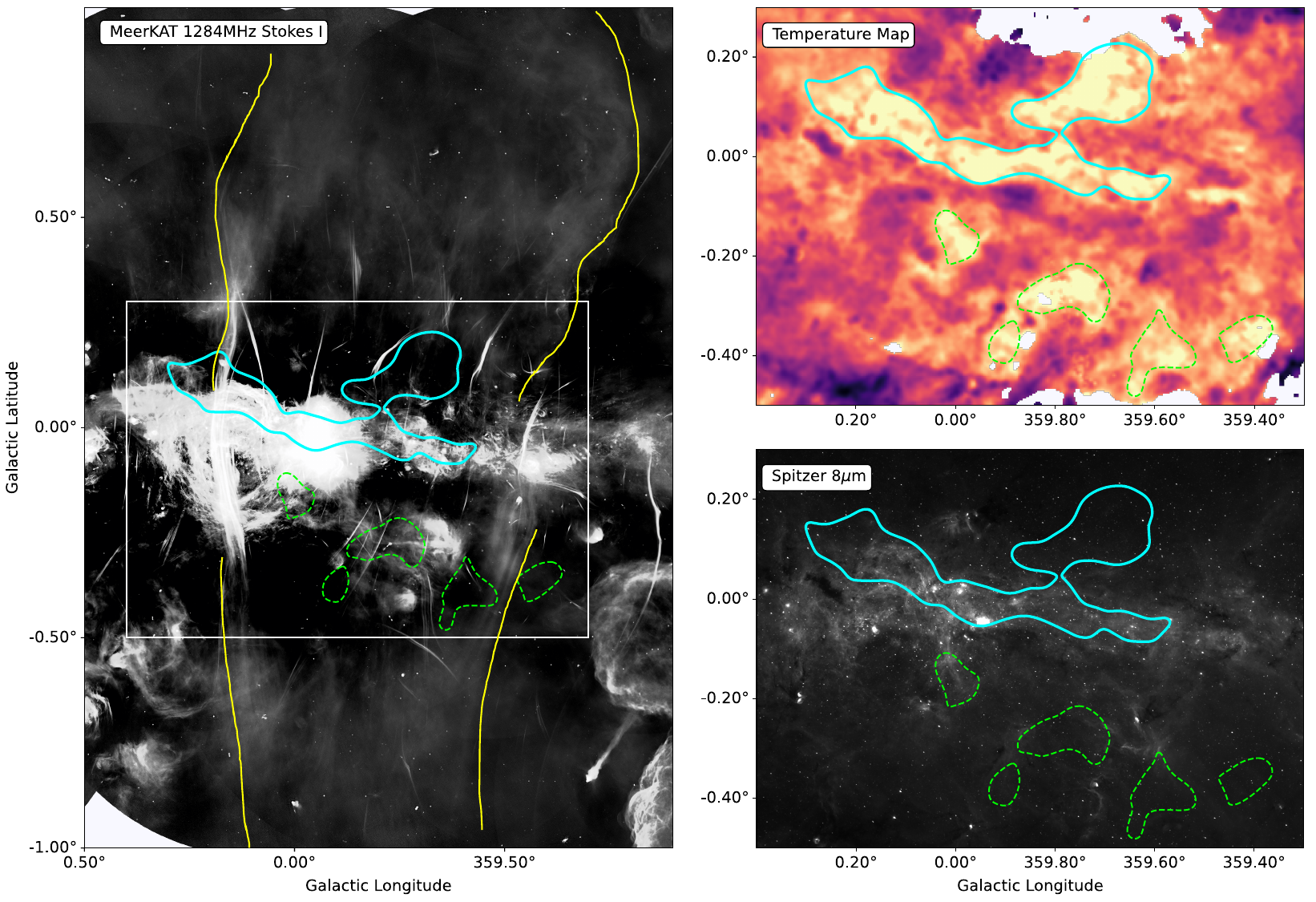}
\caption{In the CMZ, we find a ridge of warm dust (\S \ref{sec:ridge}) that could potentially trace the base of the northern Galactic outflow seen with MEERKAT (see \S \ref{sec:ridge}). The \textbf{left} panel shows MeerKAT 1284 MHz Stokes I continuum emission with approximate outlines of the Galactic-scale outflows identified in \citet{Heywood2022} in yellow. The white box indicates the zoom region of the right two plots. The \textbf{right} panel shows both the dust temperature (\textit{top}), derived in this work, and Spitzer 8\micron~emission({\textit{bottom}}) from \citet{Benjamin2003}. The solid cyan contour on all images shows a dust temperature above 29 K, highlighting the `warm dust ridge,' while the dashed green contours show other warm spots (T $> 29$ K).}
\label{fig:meerkat}
\end{figure*}

\subsection{A Ridge of Warm Dust}
\label{sec:ridge}
We find a ridge of relatively warm dust ($>$ 29 K) which runs almost parallel to the Galactic Plane, just above SgrA* from about $\ell = 0.3$\deg~to $\ell =-0.4$\deg (see Figure \ref{fig:meerkat}, solid blue contour). We identified this region based on its striking appearance in the temperature map. The warm dust ridge blue contour is drawn at a temperature of 29 K on a map smoothed with Gaussian kernel of 5.6 pixels which best described the overall shape of the ridge. We also show other regions of warm dust (T $> 29$ K) in the central area with green dashed contours. These contours do not form a clear ridge, but could plausibly be associated with the Southern outflow lobe, though the connection is even more tenuous.
This warm, dust ridge appears to be spatially related to the polar arc identified in \citet{Bally1988} or Arm III in \citet{Sofue1995}. In \citet{Henshaw2016a}, this region is shown in a green contour in Figures 9 and 11 and is also discussed in \citet{Kruijssen2015}.

In this section, we investigate the plausibility of initial hypotheses for the cause of this warm dust ridge, which are: 1) it is the base of the large outflows identified in MeerKAT \citep{Heywood2022}, 2) it is associated with massive stars \citep[as suggested in e.g.][]{Hankins2019,Hankins2020}, or 3) where the column density is low, the dust temperature is high, and this is just a coincidence of low column density. We also note that \citet{Anderson2024} suggest that part of this Galactic Center Lobe may actually just be a foreground HII region.

In order to investigate the first hypothesis, in Figure \ref{fig:meerkat} we compare the morphology of the warm dust ridge (cyan contour) with the MeerKAT data. The warm dust ridge is largely within the bounds of the MeerKAT-identified outflows and perpendicular to these outflow walls, forming a relatively long, thin ridge that is plausibly at the Northern base of this outflow. It extends into the outflow cavity in a bubble of continued warm dust. Based on the morphology of the ridge, the first hypothesis is deemed plausible but not confirmed. The Southern lobe may be associated with the other warm spots, delineated with green dashed contours in Figure \ref{fig:meerkat}, though the connection is speculative.

In the second hypothesis, the ridge of warm dust is thought to be due to heating from young, still-forming or just formed massive stars that follow the morphology of this structure. Detailed maps, including some with temperature, from \citet{Hankins2019, Hankins2020} as well as from the GLIMPSE survey \citep{Benjamin2003} show mid-IR emission from many young, massive star-forming regions in the CMZ. While there are quite a few located in the central region of the warm dust ridge just above SgrA*, otherwise there seems to be no special association of the young, forming massive star population with the morphology of the dust ridge or other warm spots. In Figure \ref{fig:meerkat} we show the warm dust contours on top of the Spitzer 8 \micron~data from the GLIMPSE survey \citep{Benjamin2003}. The central bright massive star-forming regions are clearly seen at 8 \micron, but otherwise, a visual inspection shows no trend with the morphology of the warm dust contours. The same can be seen in the SOFIA data maps from \citet{Hankins2019, Hankins2020}. We therefore conclude that it is unlikely that this dust ridge is solely caused by heating from young, massive star-forming regions in the CMZ.

The third hypothesis is that since dust temperature and column density are classically anti-correlated, perhaps the ridge of warm dust is simply tracing a low column density region. While it is certainly true that the warm dust ridge is associated with a lower column density than the central ring of the CMZ, it is far from the only region in the CMZ at a lower density. We investigate this hypothesis in Figure \ref{fig:nh2_temp_ridge} which shows a 2-D histogram of the column density vs. dust temperature for all points in the CMZ (in the inner 7\deg~map as described in Section \ref{sec:inner40}) in gray and for the warm, dust ridge in purple. We see this classic anti-correlation between column density and dust temperature and find that the warm, dust ridge is simply a high temperature portion of the material, at a range of column densities. Additionally, in Figure \ref{fig:3color_nh2_temp}, while the warm dust ridge appears as completely average in terms of its column density (middle panel), it is quite striking in the dust temperature map (bottom panel). Therefore, we see no evidence to support that the warm dust ridge is a temperature enhancement determined solely by its relatively low column density.

In conclusion, we find that the warm, dust ridge can be plausibly placed at the base of the MeerKAT-identified outflows, however, this association is still speculative. Additional investigation is required to further explore this suggestion. It is worth noting that this warm dust ridge is also seen in AzTEC maps \citep{Tang2021a, Tang2021b} as well as previous Herschel dust temperature maps \citep[e.g.][]{Molinari2011}.

\begin{figure*}
\includegraphics[width=1\textwidth]{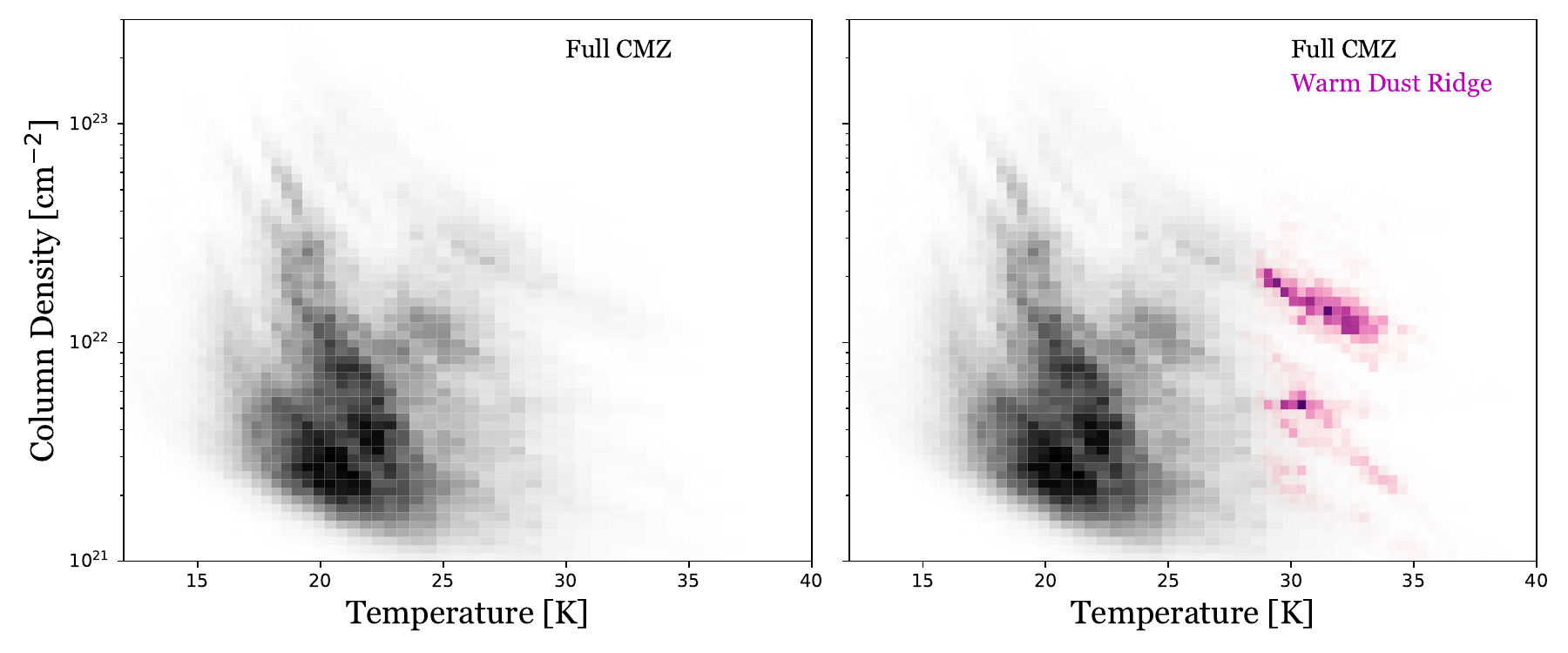}
\caption{A 2D histogram of the column densities and temperatures from this work over the entire CMZ indicates an anti-correlation between the two, with points associated with the warm dust ridge at the high temperature, low column density tail of the overall population. On the \textit{left} is a 2D histogram of column density and temperature measurements across the entire CMZ (within source masks). The highest column densities are associated with the lowest temperatures and vice versa. On the \textit{right} we overplot the 2D histogram of points associated with the warm dust ridge and find that this region is at the high temperature, low column density tail of the overall population.}
\label{fig:nh2_temp_ridge}
\end{figure*}

\subsection{Comparison with Previous Work}
\label{sec:previous}

\begin{figure*}
\subfigure{
    \includegraphics[width=0.48\textwidth]{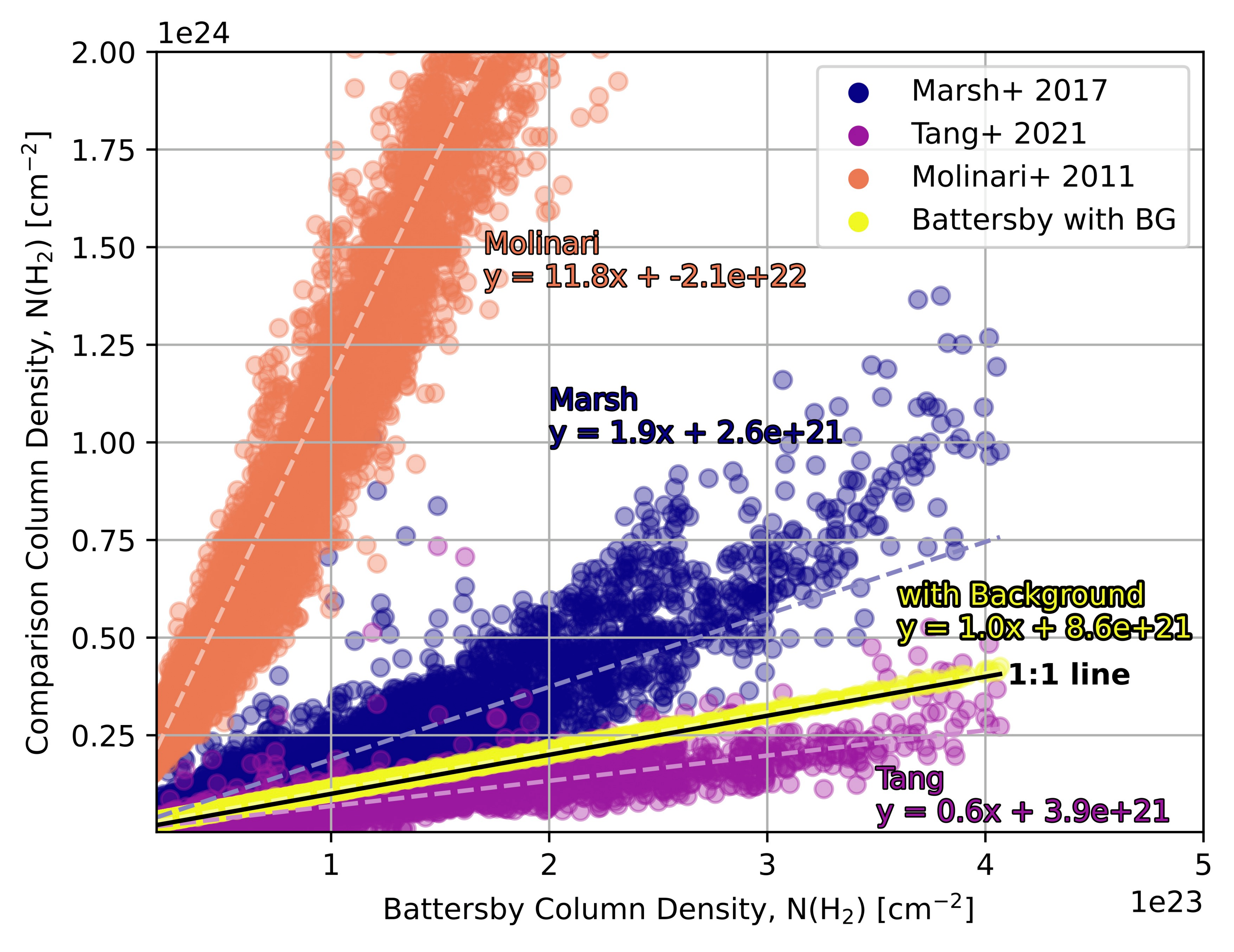}}
    \subfigure{
     \includegraphics[width=0.48\textwidth]{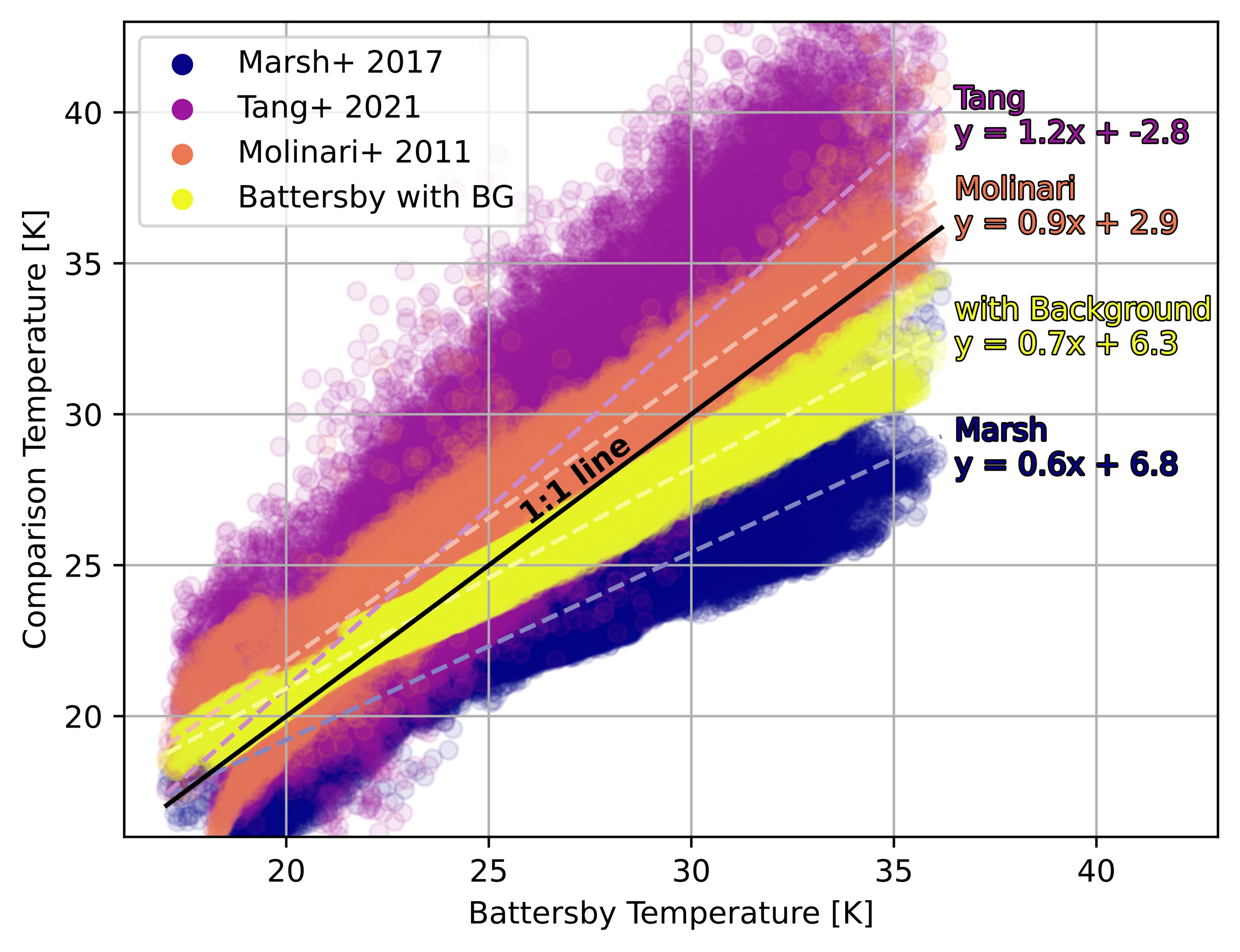}}
    \caption{A quantitative comparison of column densities and temperatures from the literature with this work shows a decent amount of scatter, but an overall good agreement, with the notable exception of the \citet{Molinari2011a} column densities. The \textit{left plot} shows the column densities computed in this work on the x-axis versus the column densities in \citet{Marsh2017} in dark blue, \citet{Tang2021a, Tang2021b}  in purple, \citet{Molinari2011a} in orange, and the column densities without background subtraction in yellow. Results of linear fits are shown in the dashed lines and expression of the same color. The \textit{right plot} shows the comparison between the dust temperatures for all regions with N(H$_2$) $> 10^{23}$ cm$^{-2}$ from this work on the x-axis, and other regions in colors that match the left plot. Linear fits are again displayed as dashed lines and expressions of the same color.}
    \label{fig:lit_comp}
\end{figure*}

\subsubsection{CMZ Column Densities and Dust Temperatures}

We compare our dust temperature and column density measurements with others in the literature, \citet{Marsh2017}, \citet{Tang2021a, Tang2021b}, \citet{Molinari2011a}, which are based on the same dataset but use a variety of techniques. We present a quantitative comparison in Figure \ref{fig:lit_comp}. The left panel shows the comparison between the column densities, N(H$_2$), along with linear fits and the right panel shows the comparison with dust temperatures along with linear fits. These plots show a decent amount of scatter, but overall good agreement between both dust temperatures and column densities. The notable exception is a large disagreement with the \citet{Molinari2011a} column densities.

To create Figure \ref{fig:lit_comp} we retrieved published datasets from the hyperlinks indicated in publications for \citet{Marsh2017} and \citet{Tang2021a, Tang2021b} and via private communication for \citet{Molinari2011a}. We used the package {\sc{astropy reproject}} to reproject all of the datasets onto the smallest footprint, which was \citet{Tang2021a,Tang2021b}. We convert all of the data to the same units, remove NaNs, and then plot the pixel-by-pixel comparison between each dataset and the measurements in this work. Since the pixel size is smaller than the beam, these points are not all independent. We use the package {\sc{numpy polyfit}} to perform simple linear fits of each dataset relative to the measurements in this work. For both measurements, we only fit a line to points with a column density N(H$_2$) $> 10^{22}$ cm$^{-2}$ since this is approximately the level of the background, making data below this point less useful in comparison. 

Each of these measurements have advantages and disadvantages. Our work uniquely and uniformly provides the column density and dust temperature measurements across the entire inner 40\deg~longitude of the Galaxy, therefore we do not include higher resolution data available only in the CMZ. One advantage of this approach is a comprehensive fore/background subtraction process, applied automatically over the entire inner 40\deg. This fore/background subtraction process was important for our purposes since the following papers in this series are focused on the hierarchical cloud-scale and CMZ-wide properties of the dense gas. However,  we enthusiastically refer the reader to other works \citep[particularly][]{Tang2021a, Tang2021b, Marsh2017} for higher resolution column density and temperature maps in just the CMZ.

Here is a summary of the datasets to which we compare. \citet{Molinari2011a} performed simple, non-fore/background-subtracted SED fits to the Herschel data as part of the initial data release. They utilized wavebands from $70-350$\micron~(as opposed to our $160-500$\micron~approach), yielding a higher spatial resolution, but with possible complications from the ambiguity of the 70\micron~data point which can be contaminated by the emission from very small dust grains, \citep[e.g.][]{Desert1990, Compiegne2010}. We also compare with \citet{Marsh2017} who calculated dust temperatures and column densities from the Herschel data using the `PPMAP' technique, which aims to deconstruct different temperature components along the line-of-sight to achieve an overall higher spatial resolution. Finally, we compare with \citet{Tang2021a, Tang2021b} who combined the Herschel data analyzed in this work with extant Bolocam \citep{Aguirre2011, Ginsburg2013} and Planck data \citep{PlanckCollaboration2011}, and new data from AzTEC on the LMT \citep{Tang2021b}, to produce full dust SEDs across the Galactic Center region (1.6 $\times$ 1.1 deg$^2$). We also refer the reader to \citet{Barnes2017} who utilize the same data and fore/background subtraction approach as this work, but extend the analysis to include both a cold dust component (as we analyze here) as well as a warm dust component using Spitzer data. We further synthesize with \citet{Barnes2017} in Paper II of this series.

As seen in the right panel of Figure \ref{fig:lit_comp}, dust temperatures show overall good agreement, including of the warm, dust ridge (Section \ref{sec:ridge}). In addition to the linear fits shown on the plot, we find that the overall ratio between each literature temperatures and our temperatures (e.g. Tang temp / Battersby temp = ratio) is as follows: the \citet{Tang2021a, Tang2021b} ratio has a median value of 1.07 and a standard deviation of 0.08, the \citet{Molinari2011a} ratio is 1.06 $\pm$ 0.03, the with background temperatures have a ratio of 0.98 $\pm$ 0.03, and the \citet{Marsh2017} ratio is 0.87 $\pm$ 0.04. We note that the agreement with \citet{Molinari2011a} and  \citet{Tang2021a, Tang2021b} is especially good in regions of high column density, but in low column density regions, the variation can be higher, with the \citet{Tang2021a, Tang2021b} temperatures being systematically higher overall. The overall trends in temperature are consistent among our maps and the PPMAP results. However, the PPMAP technique finds a uniformly lower dust temperature. While this is likely partly due to the higher spatial resolution in the PPMAP images, this cannot be the full explanation, since we would expect not only deeper troughs but also higher peaks if this were purely a resolution effect. 

As seen in the left panel of Figure \ref{fig:lit_comp}, the column densities derived show overall reasonable agreement with previous works in the literature \citep{Marsh2017, Tang2021a, Tang2021b}. In addition to the linear fits shown on the plot, we find that the overall ratio between each literature column density and our column density (e.g. Tang column density / Battersby column density = ratio) is as follows: the \citet{Tang2021a, Tang2021b} ratio has a median value of 0.84 and a standard deviation of 0.18, the \citet{Molinari2011a} ratio is 11.4 $\pm$ 1.4, the with background column densities have a ratio of 1.41 $\pm$ 0.21, and the \citet{Marsh2017} ratio is 2.1 $\pm$ 0.43.
The column densities we report, are significantly lower than those from \citet{Molinari2011a}. Some difference is expected since that work did not subtract a fore/background which can account for up to a factor of two and they also used a relationship between $\tau_{dust}$ and $N_H$ that is calibrated at high galactic latitudes which could be a factor of 2-4 higher in the plane \citep{Molinari2011a}. \citet{Ponti15} also find that column densities derived by Molinari et al. (2011) are larger than their constraints from X-ray modeling by a factor of 5-10. Our values agree with reasonable certainty with the \citet{Tang2021a, Tang2021b} but our values are systematically lower than those from \citet{Marsh2017} by about a factor of 2. This is likely due to a combination of factors. For one, our resolution is lower, so high column density regions would be smeared out in our maps and have a higher intrinsic column density in the \citet{Marsh2017} approach. Secondly, our careful fore/background subtraction technique removes some contaminating emission which leads to a lower overall value of the column density. Finally, the PPMAP temperatures are lower, which would lead to a higher measured column density for similar flux values. Additionally, each paper used slightly different dust opacities, which can explain some amount of discrepancy in the resulting column densities. As described in more detail in Section \ref{sec:uncertainties}, these discrepancies (with the exception of the \citet{Molinari2011a} column densities) are in line with the intrinsic uncertainty expected in these measurements of about a factor of two in the column densities.

\subsubsection{CMZ Gas Temperatures}

Previous measurements of the gas temperature in the CMZ, obtained using spectral line ratios, have reported gas temperatures that are uniformly higher than measured dust temperatures \citep[e.g.][]{Henshaw2023, Krieger2017, Ginsburg2016, Immer2016, Ott2014, Mills2013, Ao2013, Huettemeister1993, Gusten1981}. This is also seen in simulations of the CMZ \citep{Clark2013} and is a consequence of the the strong heating of the CMZ gas by radiation, cosmic rays and turbulence, which delays the coupling between the gas and dust temperature until the density reaches n $\sim$ $10^7$ cm$^{-3}$. The trend of CMZ gas temperature measurements being consistently higher (generally $T>60-200$ K) than dust temperatures is seen on both large ($\sim$ 1 pc) and small ($\le$ 0.1 pc) scales. Our results are in agreement with this well-documented trend \citep[e.g.][]{Henshaw2023, Krieger2017}. However, the spatial trends seen in dust temperature that are discussed in this paper have not been yet reported in maps of the gas temperature, a subject worthy of future investigation.

\section{Conclusions}
\label{sec:conclusions}

We have presented an overview of the far-IR dust continuum perspective of the CMZ of our Galaxy. Using a careful fore/background-subtraction approach and single-temperature modified blackbody fits to Herschel data, we have derived a complete fore/background-subtracted column density and dust temperature map of the inner 40\deg~of the Galaxy. These maps match the angular resolution of Herschel with an approximately 36\arcsec~beam. These data generally agree with previous measurements of the column density and dust temperature in overlapping regions. These maps, and all other products from this work, are made publicly available on the Harvard dataverse (\href{https://dataverse.harvard.edu/dataverse/3D_CMZ}{https://dataverse.harvard.edu/dataverse/3D\textunderscore CMZ}) as described in Section \ref{sec:release}. Below, we summarize key conclusions drawn from this work:

\begin{itemize}

\item The CMZ clearly stands out from the rest of the inner 40\deg~of the Galaxy as a large, contiguous region of high column density in the Galaxy's Center with an extent of approximately 1.8\deg $> \ell >$ -1.3\deg in Galactic longitude. 

\item We find typical CMZ cloud column densities to be about 10$^{23}$ cm$^{-2}$, with the highest peak at SgrB2 with $N$(H$_2$) = 2 $\times 10^{24}$ cm$^{-2}$ and several un-recovered saturated pixels likely above this value. 

\item The range of CMZ dust temperatures is relatively small, from about 12 - 35 K. The highest dust temperatures are towards the inner 100 pc of the CMZ, followed by the SgrB2 region, with the lowest dust temperatures outside the inner 100 pc.

\item We report a total CMZ dense gas mass of M$_{\rm{dense}}$ $\sim 2\substack{+2 \\ -1} \times 10^7$ \Msun. We confirm previous reports of a highly asymmetric dense gas distribution in the CMZ, with about 70-75\% of dense gas at positive longitudes, and a slight latitude asymmetry as well, with the majority of dense gas at b $<$ 0\deg.

\item We identify a ridge of warm dust in the inner CMZ from about 0.3\deg $< \ell < -0.4$\deg~and 0.2\deg $< b < -0.05$\deg~that potentially traces the base of the northern Galactic outflow seen with MEERKAT.

\end{itemize}

\acknowledgments

C.\ Battersby  gratefully  acknowledges  funding  from  National  Science  Foundation(NSF)  under  Award  Nos. 1816715, 2108938, 2206510, and CAREER 2145689, as well as from the National Aeronautics and Space Administration through the Astrophysics Data Analysis Program under Award ``3-D MC: Mapping Circumnuclear Molecular Clouds from X-ray to Radio,” Grant No. 80NSSC22K1125.
D. Walker and D. Lipman gratefully acknowledge funding from the NSF under Award No. 1816715 and D. Lipman also acknowledges funding from the National Science Foundation under Award Nos. 2108938, and CAREER 2145689, as well as from the NASA Connecticut Space Grant Consortium under PTE Federal Award No: 80NSSC20M0129.
A. Ginsburg acknowledges funding from NSF awards AST 2008101 and 2206511 and CAREER 2142300.
J.\ D.\ Henshaw gratefully acknowledges financial support from the Royal Society (University Research Fellowship; URF/R1/221620). 
RSK and SCOG acknowledge financial support from the European Research Council via the ERC Synergy Grant ``ECOGAL'' (project ID 855130),  from the German Excellence Strategy via the Heidelberg Cluster of Excellence (EXC 2181 - 390900948) ``STRUCTURES'', and from the German Ministry for Economic Affairs and Climate Action in project ``MAINN'' (funding ID 50OO2206). 
E.A.C.\ Mills  gratefully  acknowledges  funding  from the National  Science  Foundation  under  Award  Nos. 1813765, 2115428, 2206509, and CAREER 2339670.
M. C. Sormani acknowledges financial support from the European Research Council under the ERC Starting Grant ``GalFlow'' (grant 101116226).
Q. Zhang acknowledges the support from the National Science Foundation under Award No. 2206512.

\software{This work made use of Astropy:\footnote{http://www.astropy.org} a community-developed core Python package and an ecosystem of tools and resources for astronomy \citep{astropy:2013, astropy:2018, astropy:2022}.
This research made use of astrodendro, a Python package to compute dendrograms of Astronomical data (\href{http://www.dendrograms.org/}{http://www.dendrograms.org/}) as well as SAO Image ds9 \citep{Joye2003}.
This research made use of Montage. It is funded by the National Science Foundation under Grant Number ACI-1440620, and was previously funded by the National Aeronautics and Space Administration's Earth Science Technology Office, Computation Technologies Project, under Cooperative Agreement Number NCC5-626 between NASA and the California Institute of Technology.
This research has made use of NASA's Astrophysics Data System Bibliographic Services.}

\clearpage

\appendix
\section{Appendix: Data Release}
\label{sec:appendix_data}
All data analysis products from this work are freely available on the Harvard Dataverse here: \href{https://dataverse.harvard.edu/dataverse/3D_CMZ}{https://dataverse.harvard.edu/dataverse/3D\textunderscore CMZ}. The dataset for this paper is found in the Papers I and II repository: \href{https://doi.org/10.7910/DVN/7DOJG5}{https://doi.org/10.7910/DVN/7DOJG5} \citep{Battersby_3DCMZII_data}.  We describe here the data products, and this information is also available in the Dataverse metadata.

The data products are all normalized to the same resolution and pixel grid matching our lowest resolution at 36\arcsec, and are provided for both the inner 7\deg~of the Galaxy as well as the full inner 40\deg~of the Galaxy. For each region (inner 7\deg~ and inner 40\deg~), we provide:
\begin{itemize}
\item a label map showing the location of the source masks (e.g. higalcmz\_column\_density\_inner40deg\_label.fits),
\item a column density map with the fore/background-subtracted column density inside the source masks and the smoothed fore/background column density outside the source masks (e.g. higalcmz\_column\_density\_inner40deg.fits),
\item a column density map with only the fore/background-subtracted column density inside the source masks and NaNs outside the source mask (e.g. higalcmz\_column\_density\_source\_only\_inner40deg.fits),
\item a column density map without any fore/background subtraction (e.g. higalcmz\_WITH\_BG\_column\_density\_inner40deg.fits),
\item a temperature map with the fore/background-subtracted temperature inside the source masks and the smoothed fore/background temperature outside the source masks (e.g. higalcmz\_temperature\_inner40deg.fits),
\item a temperature map with the only the fore/background-subtracted temperature inside the source masks and NaNs outside the source mask (e.g. higalcmz\_temperature\_source\_only\_inner40deg.fits), and
\item a temperature map without any fore/background subtraction (e.g. higalcmz\_WITH\_BG\_temperature\_inner40deg.fits).
\end{itemize}

\bibliography{refs_3dcmz.bib}{}

\end{document}